\documentclass[aps,prd,preprint,groupedaddress]{revtex4}
\usepackage{epsf,epsfig}
\newcommand{\Exp}[1]{{\mathrm{e}}^{#1}}
\newcommand{\be}{\begin{equation}}
\newcommand{\ee}{\end{equation}}
\newcommand{\bea}{\begin{eqnarray}}
\newcommand{\eea}{\end{eqnarray}}
\newcommand{\LL}{{\mathcal L}}
\newcommand{\pa}{\partial}
\newcommand{\pam}{{\partial_\mu}}

\newcommand{\ba}{\begin{array}}
\newcommand{\ea}{\end{array}}
\newcommand{\lp}{\left(}
\newcommand{\rp}{\right)}
\newcommand{\Bv}{\displaystyle{\biggl\vert}}
\newcommand{\lac}{\left\{}

\renewcommand{\a}{\alpha}

\newcommand{\g}{\gamma}
\renewcommand{\d}{\delta}

\newcommand{\om}{\omega}

\newcommand{\ep}{\epsilon}

\begin{document}

                
\title{Particle Creation from Q-Balls.}


\author{Stephen S. Clark}
\affiliation{Institut de th\'eorie des ph\'enom\` enes physiques, Ecole Polytechnique F\'ed\'erale de Lausanne, CH-1015 Lausanne, Switzerland.}


\date{\today}

\begin{abstract}
Non topological solitons, Q-balls can arise in many particle theories with $U(1)$ global symmetries. As was shown by
Cohen et al. \cite{Qballscohen}, if the corresponding scalar field couples to massless fermions, large Q-balls are unstable and 
evaporate, producing a fermion flux proportional to the Q-ball's surface. In this paper we analyse Q-ball instabilities as a
function of Q-ball size ans fermion mass. In particular, we construct an exact quantum-mechanical description of the evaporating Q-ball.
This new construction provides an alternative method to compute Q-Ball's evaporation rates. We shall also find the new expression for
the upper bound on evaporation as a function of the produced fermion mass and study the effects of Q ball's size on particle production.

\end{abstract}


\maketitle

\section{Introduction}
A scalar field theory with an unbroken continuous global symmetry admits a remarkable class of solutions, non-topological solitons or
Q-Balls. These solutions are spherically symmetric non-dissipative solutions to the classical field equations \cite{Qballscoleman,Qballscohen,Qballsnew}. 
In a certain way they can be viewed as a sort of Bose-Einstein condensate of ``classical'' scalars. 
The construction of these solutions uses the fact that they are absolute minima of the energy for a fixed value of the conserved $U(1)$-charge
Q. So in the sector of fixed charge the Q-Ball solution is the ground state and all its stability properties are due to charge conservation. An important amount of
work has been done on Q-Ball dynamics and on their stability versus decay into scalars \cite{Qballscoleman,axenides}. Apart some existence theorems that depend
on the type of symmetry and the potentials involved \cite{Qballsnew}, the stability of Q-Balls is due to the fact that their mass is smaller than the mass
of a collection of scalars. 

In realistic theories the scalar field has a coupling with fermionic fields. The addition of this coupling modifies the criterions of Q-Ball stability since
they can now produce  fermions. This fact will have an important interest for cosmology since Q-Balls can play the role of dark matter 
\cite{Kusenko:1997si,Kusenko:2004yw}. Particle production from the Q-Ball will reduce its charge Q and at a certain point the Q-Ball will become unstable
versus decay into scalars to finally disappear. This problem has been considered in \cite{Qballscohen,Qballsnew} for the production of massless
fermions by a large Q-ball. The method used was to construct the quantum field as a superposition, with operator valued coefficients, of the classical solutions. 
In most cases we can express the general solution as a superposition of partial waves. The next step will be to use the asymptotic behaviour of the fields, 
considering the far past and the far future where the fields can be identified to free ones. In most configurations the fields in the far past
and in the far future are free fields and the relation, the $S$-matrix, linking them together contains all the information needed to
answer the question of particle production. The problem we have here is that the Q-ball is a time-dependent configuration, so we need to be careful
when we identify the asymptotic states to free (static) ones. We must make sure that the identification is made before the interaction is turned on and
after it is turned off.

To avoid the problems linked to the time dependence of the Q-ball we shall propose an alternative method. This uses
Heisenberg's picture of quantum mechanics. We shall construct the time independent state representing particle production, it is done
by solving the condition that no fluxes are coming from infinity (no particles are moving towards the Q-ball). 
We can then build the Heisenberg field operator containing all the
relevant information and time dependence. Particle creation is then computed by using the number operator, but any other operator valued
quantity can be calculated. The use of this method needs no limit calculations on the Q-ball's size, so it can be used to study small
Q-ball as well. The difficulty now lies in solving the production condition. This
condition can be solved considering asymptotics of the fields far away from the Q-ball. 

The major difference between the standard construction and this one lies in the kinematical conditions used. The standard S-matrix based
method will solve matching conditions to compute all the reflection and transmission amplitudes of an incident wave. The approach used here is different in the
sense that we construct a state having no incoming wave, making the computation of scattered amplitudes useless. Using this construction
implies that all the particles are created inside the Q-ball, since they all move away from the Q-ball.

The other important fact to investigate is the production of massive fermions. Can a Q-Ball produce any type of fermions, and is the fermion
mass relevant for the Q-Ball's life time? To answer this question we can use both pictures, the problem is now that solutions to massive
field equations have twice as many degrees of freedom. Even if the construction of the Heisenberg field operator is possible and not very difficult,
the resolution of the particle production condition is a complex task to achieve, so we shall use the $S$-matrix picture. The partial
wave expansion of the solution can easily be obtained when working with one space dimension. In fact using this simple picture allowed us to obtain
analytical results describing a fermion field being scattered by a Q-ball. Computing the evaporation rate as a function of the produced fermion
mass will lead to a new definition for the absolute upper bound of evaporation rate. An other important question we can ask is the role of the
Q ball's size on the different particle creation regimes. To answer this question we shall study both very big and small Q balls. 

The paper is organised as follows. We first give a review of the simplest 3 dimensional Q-Ball model to build up its basic properties and we then reduce 
it to 1 space dimension. We then consider interaction of Q-Balls with massless fermions
and construct a solution where there are only fermions moving away from the Q-Ball, this construction will give us the evaporation rate of Q-Balls for
production of massless fermions. We then consider production of massive fermions, using the incident wave outside the Q-Ball. This requires the knowledge
of the reflection and transmission amplitudes on the Q-Ball's surface.  Finally we extend our results to $3\oplus1$ dimensions.
\section{A simple Q-Ball model}
We review here the basic properties of a 3-dimensional Q-Ball using the simplest possible model. As we mentioned in the introduction, the Q-Ball is the
ground state of a scalar theory containing a global symmetry. 
We can now build the simplest model in $3\oplus1$ dimensions having a Q-Ball solution: it is a $SO(2)$ invariant theory 
of two real scalar fields (in fact it is the $U(1)$ theory of one complex scalar field) \cite{Qballscoleman}. We start by writing down the Lagrangian and the
equations of motion for the scalar field, to obtain the conserved charge and current.
The Lagrangian of the scalar sector is given by :
\begin{eqnarray}
\LL=\pam\phi^\star\pa^\mu\phi-U(|\phi|).
\end{eqnarray}
The $U(1)$ symmetry is
\begin{eqnarray}
\phi\rightarrow\Exp{i\alpha}\phi. \nonumber
\end{eqnarray}
The conserved current is 
\begin{eqnarray}
j_\mu=i(\phi^\star\pam\phi-(\pam\phi^\star)\phi),
\end{eqnarray}
and the conserved charge is
\begin{eqnarray}
Q=\int d^3xj_0.
\end{eqnarray}
We build a solution with the minimal energy : if $U(0)=0$ is the absolute minimum of the potential, $\phi=0$ is the 
ground state and the $U(1)$ symmetry is unbroken. It was shown in \cite{Qballscoleman} that new particles 
(Q-Balls) appear in the spectrum, if the potential is such that the minimum of $\frac{U}{|\phi|^2}$ is at some value 
$\phi_0\neq0$.
\begin{eqnarray}
Min[2U/|\phi|^2]=2U_0/|\phi_0|^2<\mu^2=U''(0).
\end{eqnarray}
The charge and energy of a given $\phi$ field configuration are :
\begin{eqnarray}
\ba{c}Q=\frac{1}{2i}\int(-\pa_t\phi^\star\phi+c.c.)d^3x,  \\
E=\int\left[\frac{1}{2}|\dot{\phi}|^2+\frac{1}{2}|\nabla\phi|^2+U(\phi)\right]d^3x.\ea\label{Qdef}
\end{eqnarray}
The Q-Ball solution is a solution with minimum energy for a fixed charge, we thus introduce the following Lagrange
multiplier
\begin{eqnarray}
\varepsilon_\om=E+\om[Q-\frac{1}{2i}\int(\phi^\star\pa_t\phi+c.c.)d^3x].
\end{eqnarray}
Minimising this functional with the standard  Q-Ball ansatz :
\begin{eqnarray}
\phi=\phi(\vec{x})\Exp{i\om t},
\end{eqnarray}
where $\phi(r)$ is a monotonically decreasing function of distance to the origin, and zero at infinity. 
Inserting the Q-Ball ansatz in the equations of motion gives in spherical coordinates
\begin{eqnarray}
\frac{d^2\phi}{dr^2}=-\frac{2}{r}\frac{d\phi}{dr}-\om^2\phi+U'(\phi).
\end{eqnarray}
If we interpret $\phi$ as a particle position and $r$ as time this equation is similar to a Newtonian equation of motion
for a particle of unit mass subject to viscous damping moving in the potential $\frac{1}{2}\om^2\phi^2-U$.
We are searching for a solution in which the particle starts at $t=0$ at some position $\phi(0)$, at rest, 
$\frac{d\phi}{dr}=0$, and comes to rest at infinite time at $\phi=0$. Solving this problem is not difficult (see 
\cite{Qballscoleman} for details). One of the solutions can be the localised step function. Although we can  
solve exactly the equation of motion,  we will not do it in this work.

This construction is the Q-Ball we where looking for,
in the sense that it is the ground state of the theory with constant charge. We used only one field to describe it but it is in fact made of a collection of
scalars. Q-Balls rotate with constant angular velocity in internal space and are spherically symmetric in position space. 
As the charge $Q$ goes to infinity, $\om$ approaches
\begin{eqnarray}
\om_0=\sqrt{2U_0/\phi^2},
\end{eqnarray}
where $U_0$ is the value of the potential at the minimum. In this limit, $\phi$ resembles a smoothed-out step function. The
two regions ($r<R$ and $r>R$) are separated by a transition zone of thickness $\mu^{-1}$. This leads to the consideration
of two approximations the thick and thin wall regime (see \cite{Qballsnew} and \cite{kusenko} for details). 
The radius of the Q-Ball can easily be calculated using the definition of charge:
\begin{eqnarray}
Q=\frac{4}{3}\pi R^3\om_0\phi_0^2.
\end{eqnarray}
This calculation has been done  $\phi(r)=0$ if $r>R$.
All the properties of the Q-Ball are now known except the exact profile of the Q-Ball field.
We shall now build up the Q-Ball solution to our problem. The energy is given by the integral (\ref{Qdef}) and
using the previous properties and taking the limit $V\rightarrow\infty$, where $V$ is the volume of the Q-Ball, 
the energy becomes
\begin{eqnarray} 
E=\frac{1}{2}\om^2|\phi|^2V+UV.\label{en}
\end{eqnarray}
The charge becomes
\begin{eqnarray}
Q=\om_0|\phi|^2V. \label{charge}
\end{eqnarray}
We wish to minimise the energy with fixed charge. Using eq. (\ref{charge}) to eliminate $\om$, we have in the limit
of $Q\rightarrow\infty$,
\begin{eqnarray}
E=\frac{1}{2}\frac{Q}{|\phi|^2V}+UV.
\end{eqnarray}
As a function of $V$ it has its minimum at
\begin{eqnarray}
V=\frac{Q}{\sqrt{2|\phi|^2U}}.
\end{eqnarray}
Here the energy is given now by
\begin{eqnarray}
E=Q\sqrt{\frac{2U}{\phi^2}},
\end{eqnarray}
With some little modifications this construction can be adapted to any dimensions. The model we shall use in the next section is a $1\oplus1$
dimensional model. In one space dimension the Q-ball profile will be the standard step function localised in a space region of size $l$.
\section{Production of massless fermions}
The solution to the problem of particle creation by a Q-ball can be solved using two different pictures. The first one is based on the $S$-matrix
formalism, using the idea that the field is free for $t\rightarrow\pm\infty$. This construction is done by finding the solution to the equations
of motion for a fermion interacting with a Q-ball, in terms of a superposition of classical solutions. The quantisation is made by upgrading 
expansion coefficients to operators, this will give us the Heisenberg field operator. The $S$-matrix will then be constructed by identifying the fields
in the far past and in the far future to fields having the exact positive and negative frequency behaviour. This method was widely used to solve the
problem for particle creation. This method was used to compute the evaporation rate of Q-balls (\cite{Qballscohen,Qballsnew}) where the expansion
was made using rotational eigenfunctions. Once the total solution is known it is simple to build the transformation from the far past to the far future.
In the far past only the incoming wave will survive and in the far future only the outgoing ones.

The construction we are going to use here is different, we shall in the first place solve the equations of motion and obtain the Heisenberg
field operator representing a fermion interacting with a Q-ball. In one space dimension this solution will be expressed in the form,
\begin{eqnarray}
\Psi_Q=\frac{1}{\sqrt{4\pi}}\int d\ep\lp\psi_Q^+(\ep,t,z)A(\ep)+\psi_Q^-(\ep,t,z)B(\ep)\rp, \nonumber
\end{eqnarray}
where the $\psi^{\pm}_Q(\ep,t,z)$ are a basis of the solution to the Dirac equation for fermions interacting with a Q-ball of charge Q.
$A(\ep)$ and $B(\ep)$ are operators depending on energy, their anti-commutation relations are the standard ones if the $\psi$
solutions satisfy proper orthogonality conditions.
The next step we shall use is consider the space asymptotics of this solution. Far away from the Q-ball
($z=\pm\infty$ for one space dimension) the solution is the standard free field solution. This identification will give us
a relation between the solution operators $A(\ep)$, $B(\ep)$ and the free asymptotic ones $a(p)$, $b(p)$. The only
difficulty in this identification is that the quantisation of the solution was made using energy (due to the time dependence of interaction)
while the asymptotical operators depend on momentum. The next step will be to define and solve the particle production condition, saying that no particles
are moving towards the Q-ball. In terms of asymptotic operators it is 
\begin{eqnarray}
\ba{c}a_L(p)|\Psi>=b_L(p)|\Psi>=0\quad\mbox{for $p>0$, on the left} \\
a_R(p)|\Psi>=b_R(p)|\Psi>=0\quad\mbox{for $p<0$, on the right.}\ea
\end{eqnarray}
The last step of the resolution will consist in using the total Heisenberg operator $\Psi$, and the particle productive state to compute the
fermionic flux giving evaporation rate. The main idea used here is to construct a solution having no incoming wave, so we do not need
to compute any reflected or transmitted coefficients. This way all the particles come from inside the Q-ball. As mentioned before this new
construction gives a good alternative to the standard scattering kinematics. The other advantage of this picture is to allow us giving
a consistent treatment of time continuity.

In the next two subsections we shall build the solution and the relation to asymptotic operators, while in the two last subsections we shall solve
the production condition and compute the evaporation rate.  
\subsection{Solutions to the equations of motion}
Writing down the Lagrangian of a massless fermion having a Yukawa interaction with a scalar field gives in one spatial dimension, 
\begin{eqnarray}
\LL_{ferm.}=i\bar{\psi}\sigma^\mu\pam\psi+(g\phi\bar{\psi}^C\psi+h.c),
\end{eqnarray}
where the $C$ superscript indicates the charge conjugated fermion. The equations of motion and their 
solutions are fully described in literature on the subject (\cite{Qballscoleman,Qballscohen,Qballsnew}). Instead of treating separately the fermion
and the anti-fermion, we shall construct the exact global solution to this problem, this solution will be made of different parts first 
the solution inside the Q-Ball (for $z\in[-l,l]$). Equations of motion for the two components of the $\Psi$ field are : 
\begin{eqnarray}
\ba{c}(i\pa_0+i\pa_z)\psi_1-g\phi\psi_2^{\star}=0, \\
(i\pa_0-i\pa_z)\psi_2^{\star}-g\phi^\star\psi_1=0. \ea
\end{eqnarray}
and $\phi=\phi_0\Exp{-i\om_0t}$ in the zone from $-l$ to $+l$ and zero everywhere else. 
Using the ansatz :
\begin{eqnarray}
\lp \ba{c} \psi_1 \\ \psi_2^{\star} \ea \rp=\lp \ba{cc} \Exp{-i\frac{\om_0}{2}t} & 0\\
0 & \Exp{i\frac{\om_0}{2}t} \ea \rp \lp\ba{c} A  \\
B\ea\rp\Exp{-i\ep t+i(k+\frac{\om_0}{2})},\label{ansatz}
\end{eqnarray}
the equations of motion are reduced to the following $2\times2$ linear system
\begin{eqnarray}
\lp\ba{cc} k-\ep & M \\ M & -(k+\ep)\ea\rp\lp\ba{c} A \\ B \ea\rp=0 . \nonumber
\end{eqnarray}
The determinant of the system gives $k=\pm\sqrt{\ep^2-M^2}\equiv \pm k_\ep$.
Solving for the two cases  $k=+k_\ep$ and $k=-k_\ep$, we obtain the solution inside the Q-Ball:
\begin{eqnarray}
\Psi_Q=\lp\ba{c}\psi_1\\ \psi_2^\star\ea\rp=
A\lp\ba{c} 1 \\ \frac{k_\ep+\ep}{M} \ea\rp\Exp{-ik_\ep z} + B\lp\ba{c} \frac{k_\ep+\ep}{M} \\ 1 \ea\rp\Exp{ik_\ep z},
\label{inside}
\end{eqnarray}
where $M=g\phi_0$, $g$ is the coupling constant and $\phi_0$ the value of the scalar field. 
The second part is the  solution when $\phi_0=0$ (outside the Q-Ball) it is,
\begin{eqnarray}
\Psi=\lp\ba{c} \psi_1 \\ \psi_2^\star \ea \rp=
\Exp{-i\ep t}\lp\ba{c} C_1^{L,R}\Exp{i\ep z} \\ C_2^{L,R}\Exp{-i\ep z}\ea\rp, \label{coeflibre}
\end{eqnarray}
where superscripts $L,R$ indicate the left and right side of the Q-Ball. In order to solve Dirac's equation everywhere
the solution needs to be continuous in space. Space continuity gives at $z=-l$ :
\begin{eqnarray}
C_1^L=A\Exp{i(k_\ep+\ep)l}+B\alpha_\ep\Exp{-i(k_\ep-\ep)l}, \nonumber \\
C_2^L=A\alpha_\ep\Exp{i(k_\ep-\ep)l}+B\Exp{-i(k_\ep+\ep)l}, \nonumber
\end{eqnarray}
and at $z=+l$,
\begin{eqnarray}
C_1^R=A\Exp{-i(k_\ep+\ep)l}+B\alpha_\ep\Exp{+i(k_\ep-\ep)l}, \nonumber\\
C_2^R=A\alpha_\ep\Exp{-i(k_\ep-\ep)l}+B\Exp{+i(k_\ep+\ep)l}. \nonumber
\end{eqnarray}
These matching relations are used to express the solution only using the parameters coming from the inner part of the solution. This
construction will allow us to build a state where there is no incoming fermion, all the fermions are now produced inside the Q-Ball.
Putting together all these parts gives the full solution continuous in space and time :
{\small
\begin{eqnarray}
\lp\ba{c}\psi_1 \\ \psi_2^\star \ea\rp=\int d\ep\lp\ba{c}
\lp\ba{c}\Exp{i\ep l}(A\Exp{ik_\ep l}+B\a_\ep\Exp{-ik_\ep l})\Exp{-i(\ep+\frac{\om_0}{2})t}\Exp{i(\ep+\frac{\om_0}{2})z} \\
\Exp{-i\ep l}(A\a_\ep\Exp{ik_\ep l}+B\Exp{-ik_\ep l})\Exp{-i(\ep-\frac{\om_0}{2})t}
\Exp{-i(\ep-\frac{\om_0}{2})z}\ea\rp z<-l \\ \\
\lp\ba{c}(A\Exp{-ik_\ep z}+B\a_\ep\Exp{ik_\ep z})\Exp{-i(\ep+\frac{\om_0}{2})t}\Exp{i\frac{\om_0}{2}z} \\
(A\a_\ep\Exp{-ik_\ep z}+B\Exp{ik_\ep z})\Exp{-i(\ep-\frac{\om_0}{2})t}
\Exp{i\frac{\om_0}{2}z}
\ea\rp -l\geq z\geq +l \\ \\
\lp\ba{c}\Exp{-i\ep l}(A\Exp{-ik_\ep l}+B\a_\ep\Exp{ik_\ep l})\Exp{-i(\ep+\frac{\om_0}{2})t}\Exp{i(\ep+\frac{\om_0}{2})z} \\
\Exp{i\ep l}(A\a_\ep\Exp{-ik_\ep l}+B\Exp{ik_\ep l})\Exp{-i(\ep-\frac{\om_0}{2})t}
\Exp{-i(\ep-\frac{\om_0}{2})z}\ea\rp z>+l
\ea \rp, \label{solution1}
\end{eqnarray}
}
where
\bea
\a_\ep=\frac{k_\ep +\ep}{M}. \label{alpha}
\eea
A little work on the solution and on its orthogonality properties leads to a solution of the form :
\begin{eqnarray}
\Psi_Q=\frac{1}{\sqrt{4\pi}}\int d\ep\Exp{-i\ep t}\lp\psi_Q^+(\ep)A(\ep)+\psi_Q^-(\ep)B(\ep)\rp
\Exp{i\frac{\om_0}{2}z}\Omega(t), \label{solQBall} 
\end{eqnarray}
with
\begin{eqnarray}
\Omega(t)=\lp\ba{cc}\Exp{-i\frac{\om_0}{2}t} & 0 \\ 0 & \Exp{i\frac{\om_0}{2}t} \ea\rp,
\end{eqnarray}
and
{\small
\begin{eqnarray}
\psi^\pm=\lp\ba{c} \lp\ba{c}f_1^\pm(\ep,l)\Exp{i\ep z} \\
(f_2^\pm(\ep,l))^\star\Exp{-i\ep z}\ea\rp z<-l \\ \\
\frac{1}{\sqrt{N_\pm}}\lp\ba{c}(\pm\Exp{-ik_\ep z}+\a_\ep\Exp{ik_\ep z})
\\ (\pm\a_\ep\Exp{-ik_\ep z}+\Exp{ik_\ep z}) \ea\rp -l\geq z\geq +l \\
\\ \lp\ba{c} f_1^\pm(\ep,-l)\Exp{i\ep z} \\ (f_2^\pm(\ep,-l))^\star\Exp{-i\ep
z}\ea\rp z>+l \ea \rp, \label{solution3}
\end{eqnarray}}
and the functions $f^\pm_{1,2}$  having the form 
\begin{eqnarray}
\ba{c}f_1^\pm(\ep,l)=\frac{1}{\sqrt{4\pi N_\pm}}\Exp{i\ep l}(\pm\Exp{ik_\ep
l}+\a_\ep\Exp{-ik_\ep l}), \\ 
f_2^\pm(\ep,l)=\frac{1}{\sqrt{4\pi N_\pm}}\Exp{i\ep
l}(\pm\a^\star_\ep\Exp{-ik^\star_\ep l}+\Exp{ik_\ep l}).\ea
\end{eqnarray}
\begin{eqnarray}
N_\pm&=&4\pi\lp\cosh[\mathrm{Im}[k_\ep] l](1+|\a_{\ep}|^2)\right. \nonumber \\
&\pm&\left.\cos[\mathrm{Re}[k_\ep]l]\mathrm{Re}[\a_\ep]\rp,
\end{eqnarray}
Finally the time dependent matrix was introduced for simplicity, the $N_{\pm}$ are the normalisation constants. Quantisation of solution (\ref{solQBall})
was made using equal time anti-commutation relations for $\Psi$. If the $\psi^\pm$ functions satisfy 
$\int dz(\psi^{\sigma'}(\ep'))^\dagger\psi^\sigma(\ep)=\delta_{\sigma'\sigma}\delta(\ep'-\ep)$
we can show that,
\begin{eqnarray}
\{A(\ep),A^\dagger(\ep')\}&=&\int dz\int dz'{\lp\psi_Q^+(z,\ep)\rp}^\dagger{\lp\psi_Q^+(z',\ep')\rp}
\times\underbrace{\{\hat{\Psi}_Q,{\lp\hat{\Psi}_Q'\rp}^\dagger\}}_{\d(z'-z)}\nonumber \\
&=&\d(\ep'-\ep). \nonumber
\end{eqnarray}
The $\Psi_Q$ solution we obtained has now being upgraded to a Heisenberg field operator describing fermions interacting with a Q-ball.
\subsection{Relation to asymptotic operators}
First we conjugate the second component of the
above solution, in order to compare it with the standard free solution for a massless fermion in $1\oplus1$
dimensions (see \cite{abdalla} for details). We look at the asymptotic behaviour of the Q-Ball solution (\ref{solQBall}). On the left 
and right-hand side of the Q-Ball, it has to be the standard free solution since the interaction is zero outside the
Q-Ball's volume. After elimination of integrals and standard manipulations and variable changes, we obtain at 
$z\rightarrow-\infty$ :
{\small
\begin{eqnarray}
\frac{1}{\sqrt{2\pi}}\lp\ba{c}\theta(p) \\ \theta(-p)\ea\rp a(p)&+&
\lp\ba{c}\theta(-p) \\ \theta(p)\ea\rp b^{\dagger}(-p)=
\lp\ba{c}f_1^+(\ep,l)A(\ep)
\Bv_{\ep=p-\frac{\om_0}{2}} \\
f_2^+(\ep,l)A^\dagger(\ep)\Bv_{\ep=p+\frac{\om_0}{2}}\ea\rp 
+\nonumber \\
&+&\lp\ba{c}f_1^-(\ep,l)B(\ep)\Bv_{\ep=p-\frac{\om_0}{2}} \\
f_2^-(\ep,l)B^\dagger(\ep)\Bv_{\ep=p+\frac{\om_0}{2}}\ea\rp. \label{assymp}
\end{eqnarray}}
$\theta(p)$ is the heavy-side function, $\theta(p)=0$ if $p$ is negative and $\theta(p)\neq0$ when $p$ is positive.
Multiplying eq. (\ref{assymp}) by $\lp\ba{c}\theta(p) \\ \theta(-p)\ea\rp^\dagger$ and by 
$\lp\ba{c}\theta(-p) \\ \theta(p)\ea\rp^\dagger$ we obtain the two following equations :
\begin{eqnarray}
\frac{1}{\sqrt{2\pi}}a_L(p)&=&[f_1^+(\ep,l)A(\ep)+f_1^-(\ep,l)B(\ep)]\Bv_{\ep=p-\frac{\om_0}{2}}\theta(p)+ \nonumber \\
&+&[f_2^+(\ep,l)A^\dagger(\ep)+f_2^-(\ep,l)B^\dagger(\ep)]\Bv_{\ep=p+\frac{\om_0}{2}}\theta(-p) ,\nonumber \\ \label{gauche1}\\
\frac{1}{\sqrt{2\pi}}b_L^\dagger(-p)&=&[f_1^+(\ep,l)A(\ep)+f_1^-(\ep,l)B(\ep)]\Bv_{\ep=p-\frac{\om_0}{2}}\theta(-p)
+ \nonumber \\
&+&[f_2^+(\ep,l)A^\dagger(\ep)+f_2^-(\ep,l)B^\dagger(\ep)]\Bv_{\ep=p+\frac{\om_0}{2}}\theta(p) \nonumber.
\end{eqnarray}
In these two equations the $l$ subscript indicates we are on the left-hand side of the Q-Ball,  the same manipulations
on the right-hand side lead to :
\begin{eqnarray}
\frac{1}{\sqrt{2\pi}}a_R(p)&=&[f_1^+(\ep,-l)A(\ep)+f_1^-(\ep,-l)B(\ep)]\Bv_{\ep=p-\frac{\om_0}{2}}\theta(p)+ \nonumber \\
&+&[f_2^+(\ep,-l)A^\dagger(\ep)+f_2^-(\ep,-l)B^\dagger(\ep)]\Bv_{\ep=p+\frac{\om_0}{2}}\theta(-p),\nonumber \\ \label{droite2} \\
\frac{1}{\sqrt{2\pi}}b_R^\dagger(-p)&=&[f_1^+(\ep,-l)A(\ep)+f_1^-(\ep,-l)B(\ep)]\Bv_{\ep=p-\frac{\om_0}{2}}\theta(-p)
+ \nonumber \\
&+&[f_2^+(\ep,-l)A^\dagger(\ep)+f_2^-(\ep,-l)B^\dagger(\ep)]\Bv_{\ep=p+\frac{\om_0}{2}}\theta(p).\nonumber
\end{eqnarray}
Checking the anti-commutation relations of operators $a_{L,R}$ and $b_{L,R}$ is a tedious
task but using the different energy ranges and orthogonality properties of the $f_{1,2}^\pm$ it can
be done. These four equations will be the basis of the construction of particle productive state, since they give 
a relation between free operators (lower case) and solution operators (upper case). These relations will
give the solution in terms of free operators, so the next task we need to achieve is to define and construct the particle
productive state. 
\subsection{Construction of particle productive state}
As mentioned before, the construction of this quantum state $\Psi$ will be done using the fact that there are no particles moving
towards the Q-Ball. These are negative
momentum particles on the left and positive momentum particles on the right. In terms of $a_{L,R}$, 
and $b_{L,R}$ operators :
\begin{eqnarray}
\ba{c}a_L(p)|\Psi>=b_L(p)|\Psi>=0\quad\mbox{for $p>0$, on the left} \\
a_R(p)|\Psi>=b_R(p)|\Psi>=0\quad\mbox{for $p<0$, on the right.}\ea
\end{eqnarray}
This construction will lead to the opposite sign of the fermionic current on the left and on the right hand side of Q-Ball
using eqs. (\ref{gauche1}-\ref{droite2}). We then obtain four equations. For positive $p$, we have :
\begin{eqnarray}
\ba{c}(f_1^+(\ep,l)A(\ep)+f_1^-(\ep,l)B(\ep))\Bv_{\ep=p-\frac{\om_0}{2}}|\Psi>=0, \\
(f_1^+(\ep,l))^\star A^\dagger(\ep)+(f_1^-(\ep,l))^\star B^\dagger(\ep)
\Bv_{\ep=-p-\frac{\om_0}{2}}|\Psi>=0,\ea \label{relp1}
\end{eqnarray} 
and for negative $p$ 
\begin{eqnarray}
\ba{c}(f_2^+(\ep,-l)A^\dagger(\ep)+f_2^-(\ep,-l)B^\dagger(\ep))\Bv_{\ep=p+\frac{\om_0}{2}}|\Psi>=0, \\
(f_2^+(\ep,-l))^\star A(\ep)+(f_2^-(\ep,-l))^\star B(\ep))\Bv_{\ep=-p+\frac{\om_0}{2}}|\Psi>=0.\ea \label{relp2}
\end{eqnarray}
Due to the relation between $\ep$, $p$, $\frac{\om_0}{2}$ given in the subindices of eqs. (\ref{relp1}, \ref{relp2}) 
and the fact that $p$ is either positive or negative, we can 
identify three ranges for $\ep$ :
\begin{itemize}
\item For $\ep>+\frac{\om_0}{2}$ we only have the following two equations :
\begin{eqnarray}
\ba{c}(f_1^+(\ep,l)A(\ep)+f_1^-(\ep,l)B(\ep))|\Psi>=0, \\
((f_2^+(\ep,-l))^\star A(\ep)+(f_2^-(\ep,-l))^\star B(\ep))|\Psi>=0.\ea
\end{eqnarray} 
\item For the negative range $\ep<-\frac{\om_0}{2}$ we have :
\begin{eqnarray}
\ba{c}((f_1^+(\ep,l))^\star A^\dagger(\ep)+(f_1^-(\ep,l))^\star B^\dagger(\ep))|\Psi>=0, \\
(f_2^+(\ep,-l)A^\dagger(\ep)+f_2^-(\ep,-l)B^\dagger(\ep))|\Psi>=0.\ea
\end{eqnarray}
\item For the middle range $\ep\in[-\frac{\om_0}{2},+\frac{\om_0}{2}]$ we have :
\begin{eqnarray}
\ba{c}(f_1^+(\ep,l)A(\ep)+f_1^-(\ep,l)B(\ep))|\Psi>=0, \label{un}\\
(f_2^+(\ep,-l)A^\dagger(\ep)+f_2^-(\ep,-l)B^\dagger(\ep))|\Psi>=0.\ea \label{deux}
\end{eqnarray}
\end{itemize} 
The range where $|\ep|>+\frac{\om_0}{2}$ is easy to solve, since we expect the solution to be the vacuum and to 
lead to no evaporation. The determinant of the matrix :
\begin{eqnarray}
\det[\lp\ba{cc} f_1^+(\ep,l) & f_1^-(\ep,l) \\ (f_2^+(\ep,-l))^\star & (f_2^-(\ep,-l))^\star \ea \rp] =2(1-\a_\ep^2),
\end{eqnarray}
is always different from zero. The only solution for an evaporating state in this range is the 
trivial solution given by :
\begin{eqnarray}
\ba{c}A(\ep)|\Psi>=B(\ep)|\Psi>=0\quad\mbox{for}\quad\ep>\frac{\om_0}{2}, \\
A^\dagger(\ep)|\Psi>=B^\dagger(\ep)|\Psi>=0\quad\mbox{for}\quad\ep<-\frac{\om_0}{2}.\ea
\end{eqnarray}
In fact these two equations are the same, because we can always use the transformation $A(\ep)=A'(\ep)\theta(\ep)+
{B'}^\dagger(\ep)\theta(-\ep)$, all equations will have vacuum solutions. Here the anti-commutation relations are trivial 
to check because of the two different energy ranges. For the middle range $\ep\in[-\frac{\om_0}{2},+\frac{\om_0}{2}]$
things are a little more complicated, this being the range where particle production occurs as first shown in 
\cite{Qballscohen}. Taking a look at solution (\ref{solution3}) in this range, only
particles are created and changing the sign of $\om_0$ changes the particle type. We now need to check normalisation of these
new operators describing the evaporating state and their anti-commutation relations. 
Defining the evaporation operators in all the energy ranges, we have   
\begin{eqnarray}
a_e(\ep)=\lac\ba{c}A^\dagger(\ep)\quad\ep<-\frac{\om_0}{2} \\
\sqrt{8\pi}(f_1^+(\ep,l)A(\ep)+f_1^-(\ep,l)B(\ep))\quad\ep\in[-\frac{\om_0}{2},+\frac{\om_0}{2}] \\
A(\ep)\quad\ep>+\frac{\om_0}{2}\ea\right.\label{aevap} ,
\end{eqnarray}
and
\begin{eqnarray}
b_e(\ep)=\lac\ba{c}B^\dagger(\ep)\quad\ep<-\frac{\om_0}{2} \\
\sqrt{8\pi}(f_1^+(\ep,l)^\star A^\dagger(\ep)-f_1^-(\ep,l)^\star B^\dagger(\ep))\quad\ep\in[-\frac{\om_0}{2},+\frac{\om_0}{2}] \\
B(\ep)\quad\ep>+\frac{\om_0}{2}\ea\right.\label{bevap},
\end{eqnarray}
where the $\sqrt{8\pi}$ factor is the normalisation $\frac{1}{\sqrt{|f_1^+(\ep,l)|^2+|f_1^-(\ep,l)|^2}}$. The anti-commutation
relations of these operators are easy to check. They use the fact that $|f_1^\pm|^2=\frac{1}{4\pi}$. The particle production state is now fully
characterised  by the simple relation :
\begin{eqnarray}
a_e(\ep)|\Psi>=b_e(\ep)|\Psi>=0.
\end{eqnarray}
This simple relation gives the ground state for a Q-Ball producing fermions. We could now compute lots of different properties of
these fermions but we compute their number. The state defined by this relation has no incident fermion, it is exactly the state we wanted
to build. The final step will now be to compute the fermionic flux and obtain the particle production rate.
\subsection{Particle production rate} \label{evaporationrate}
The particle production rate is given by the current operator  $\vec{j}^\mu(x)=\bar{\psi}(x)\g^\mu\psi(x)$, which in our case is 
\begin{eqnarray}
\psi_1^\star\psi_1-\psi_2^\star\psi_2=\vec{j}(x)
\end{eqnarray}
that we shall apply on the evaporating state defined by the vacuum for 
$a_e$ and $b_e$ operators. First we invert the systems (\ref{aevap}) and (\ref{bevap}) to obtain :
\begin{eqnarray}
&\bullet&\ep<-\frac{\om_0}{2}\quad \lac\ba{c} A(\ep)=a_e^\dagger(\ep)\\B(\ep)=b_e^\dagger(\ep)\ea\right. \\
&\bullet&\ep>\frac{\om_0}{2}\quad \lac\ba{c} A(\ep)=a_e(\ep)\\B(\ep)=b_e(\ep)\ea\right. \\
&\bullet&\ep\in[-\frac{\om_0}{2},+\frac{\om_0}{2}]\quad \lac\ba{c}A(\ep)=\frac{1}{\sqrt{8\pi}2f_1^+(\ep,l)}(a_e(\ep)
+b_e^\dagger(\ep))\\B(\ep)=\frac{1}{\sqrt{8\pi}2f_1^-(\ep,l)}(a_e(\ep)-b_e^\dagger(\ep))\ea\right.
\end{eqnarray}
Now we can compute the first term of the current on the left hand side of the Q-Ball :
Using anti-commutation relations and the separate range of integrals and the definition of $A(\ep)$ and $B(\ep)$ in
terms of  evaporation operators $a_e(\ep)$, $b_e(\ep)$ we obtain :
\begin{eqnarray}
<0|\psi_1^\dagger\psi_1|0>&=&\int_{-\infty}^{\frac{\om_0}{2}}d\ep(|f_1^+(\ep,l)|^2<0|a_e(\ep)a_e^\dagger(\ep)|0>
+|f_1^-(\ep,l)|^2<0|b_e(\ep)b_e^\dagger(\ep)|0>) \nonumber \\
&+&\frac{1}{8\pi}\int_{-\frac{\om_0}{2}}^{+\frac{\om_0}{2}}d\ep{\Bv\frac{f_1^+(\ep,l)}{2\eta}-\frac{f_1^-(\ep,l)}{2\zeta}\Bv}^2
<0|b_e(\ep)b_e^\dagger(\ep)|0>.
\end{eqnarray}
The other term  of the current, proportional to $\psi_2^\star\psi_2$, is very similar but we need to be careful with 
the fact that $\psi_2$ is proportional to $f_2^\pm(\ep,l)$. Applying the same method 
we obtain :
\begin{eqnarray}
<0|\psi_2^\dagger\psi_2|0>&=&\frac{1}{8\pi}\int_{-\frac{\om_0}{2}}^{+\frac{\om_0}{2}}d\ep{\Bv\frac{(f_2^+(\ep,l))^\star}
{2f_1^+(\ep,l)}
+\frac{(f_2^-(\ep,l))^\star}{2f_1^-(\ep,l)}\Bv}^2<0|a_e(\ep)a_e^\dagger(\ep)|0> \\
&+&\int_{+\frac{\om_0}{2}}^{+\infty}d\ep\left({|(f_2^+(\ep,l))|}^2<0|a_e(\ep)a_e^\dagger(\ep)|0>+
{|(f_2^-(\ep,l))^\star|}^2<0|b_e(\ep)b_e^\dagger(\ep)|0>\right). \nonumber
\end{eqnarray}
It is easy to check that $|f_2(\ep,l)|^2=|f_1(\ep,l)|^2$ leading to the compensation of terms with infinite bounds.
Finally the expression for the fermionic current on the left is : 
\begin{eqnarray}
\vec{j}_L&=&\int_{-\frac{\om_0}{2}}^{+\frac{\om_0}{2}}d\ep{\Bv\frac{(f_2^+(\ep,l))^\star}{2f_1^+(\ep,l)}
+\frac{(f_2^-(\ep,l))^\star}{2f_1^-(\ep,l)}\Bv}^2 \nonumber \\
&=&\int_{-\frac{\om_0}{2}}^{+\frac{\om_0}{2}}d\ep\Bv\frac{\a_\ep\sinh[2ik_\ep l]}{\Exp{2ik_\ep l}-\a_\ep^2
\Exp{-2ik_\ep l}}\Bv^2\label{current}
\end{eqnarray}
If the real part of $k_\ep$ equals zero ($\frac{\om_0}{2}\leq M$), we use the definitions
\begin{eqnarray}
k_\ep=i\sqrt{M^2-\ep^2},\quad \a_\ep=\frac{k_\ep+\ep}{M}, \quad |\a_\ep|^2=1 \nonumber
\end{eqnarray}
and the current is then:
\begin{eqnarray}
j_L=\int_{-\frac{\om_0}{2}}^{+\frac{\om_0}{2}}d\ep\frac{\sinh^2[-2\sqrt{M^2-\ep^2}l]}
{|\Exp{-2\sqrt{M^2-\ep^2}l}-\a_\ep^2\Exp{2\sqrt{M^2-\ep^2}l}|^2}.
\end{eqnarray}
In the limit $Ml\rightarrow\infty$ we can neglect the negative exponentials and the only factor we are left with is the $\frac{1}{2}$
coming from the hyperbolic sine. The 
right hand side the current is the same except for the sign, so the total current is equal to one half. Using the continuity 
equation we can write :
\begin{eqnarray}
\frac{\pa j_0}{\pa t}+\frac{\pa j_1}{\pa z}=0\Rightarrow\frac{dQ}{dt}=\int\pa_zj(z)dz=j_L-j_R=2j_L.
\end{eqnarray}  
The current does not depend on $z$ so the value of the current is constant on both sides, so after integration of the current over $\ep$ we obtain : 
\begin{equation}
\frac{dQ}{dt}=\frac{1}{4\pi}\om_0. \label{result}
\end{equation}
This expression gives the particle production rate as a function of $\om_0$ when $\om_0$ is smaller than $M$ in the limit of big
$Ml$. It is in fact \cite{Qballscohen} an evaporation rate since it does
not depend on the Q-Ball's size. The other importance of this result is that it gives a absolute upper bound on evaporation rate.
The case when the imaginary part of $k_\ep$ is equal to zero is a bit more complicated to solve. In this case we need to explicitly compute the current
integral, so this case will be studied numerically. The value obtained for the evaporation rate in (\ref{result}) is the $1\oplus1$ equivalent of the results
found in literature on the subject.
\subsubsection{Production rate in function of size}
\begin{figure}
\begin{center}
\includegraphics{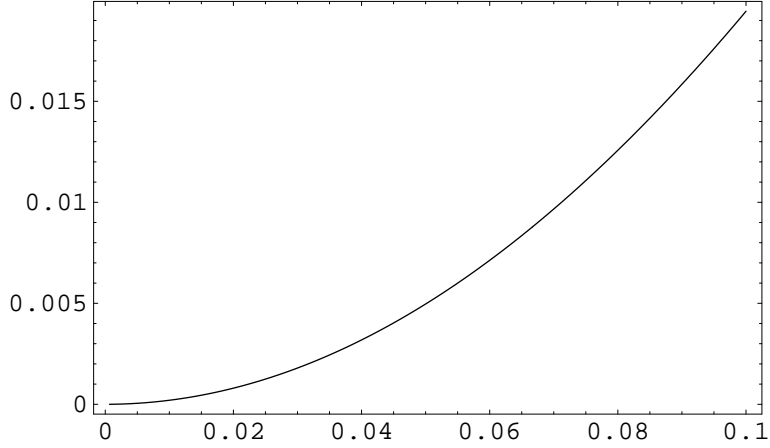}
\caption{Particle production rate for small values of $Ml$ and for a fixed value of $\frac{\om_0}{2M}=0.5$.}
\label{figl1}
\end{center}
\end{figure}

\begin{figure}
\begin{center}
\includegraphics{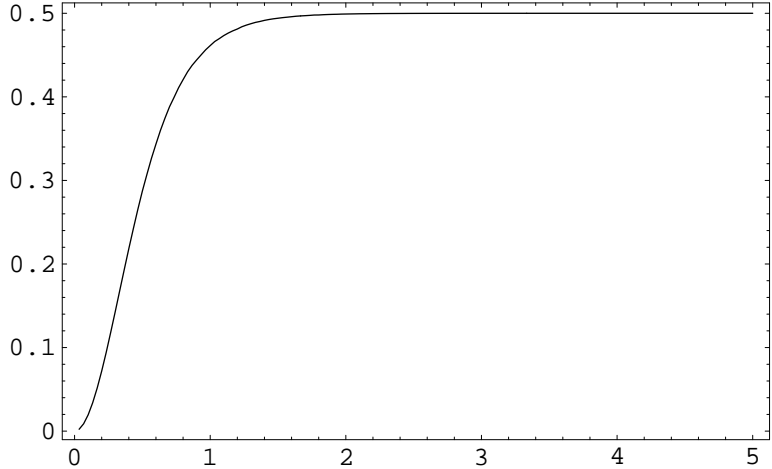}
\caption{Particle production rate in function of $Ml$ and for a fixed value of $\frac{\om_0}{2M}=0.5$.}
\label{figl2}
\end{center}
\end{figure}

We shall first consider the limit where $l$ is small. In this case we write
\begin{eqnarray}
\sinh^2[2\sqrt{M^2-\ep^2}l]&=&4(M^2-\ep^2)l^2=4(Ml)^2(1-(\frac{\ep}{M})^2) \nonumber \\
\Bv\Exp{-2\sqrt{M^2-\ep^2}l}-\a_\ep^2\Exp{2\sqrt{M^2-\ep^2}l}\Bv^2&=&\Exp{-4\sqrt{M^2-\ep^2}l}+\Exp{+4\sqrt{M^2-\ep^2}l}
-2{\mathcal Re}[\a_\ep^2] \nonumber \\
&=&2(1-\frac{2\ep-M^2}{M^2})=2(1-\frac{\ep^2}{M^2}).
\end{eqnarray}
These two terms will simplify to give after integration over $\ep$ :
\begin{eqnarray}
j_L=l^2M^2\frac{\om_0}{8\pi},
\end{eqnarray}
leading to the particle production rate:
\begin{equation}
\frac{dQ}{dt}=l^2M^2\frac{1}{4\pi}\om_0.
\end{equation}
This result enshures us the fact that when $Ml=0$, the Q ball does not exist, the evaporation rate is zero. This behaviour is shown on figure \ref{figl1}.
The next limit we shall study is the very large Q ball limit. To do so we take a look at the production rate for large values of the size parameter,
$Ml$ and observe that the production rate becomes constant for big values of the size parameter (see fig \ref{figl2}).
These considerations also stand for all the possible values of the frequency parameter the only difference is when $\frac{\om_0}{2M}$ gets bigger
the stability of the evaporation rate comes for bigger values of $Ml$.
\subsubsection{Energy flux far away form the Q ball}
The next calculation we can do is the calculation of the energy flux far away from the Q-ball. In the case where we
consider the observer very far from the Q-ball the only relevant coordinate is the distance to the Q-ball, we are in a one spacial dimension case.
The energy flux a distant observer can measure is given after normalisation by $M$,
\begin{eqnarray}
\frac{dE}{Mdtd\sigma}=\int_{-\frac{\om_0}{2M}}^{+\frac{\om_0}{2M}}d(\frac{\ep}{M})\Bv\frac{\a_{(\frac{\ep}{M})}\sinh[2i\bar{k}_{(\frac{\ep}{M})}\bar{l}]}
{\Exp{2ik_{(\frac{\ep}{M})}\bar{l}}-\bar{\a}_{(\frac{\ep}{M})}^2\Exp{-2ik_{(\frac{\ep}{M})}\bar{l}}}\Bv^2(\frac{\ep}{M})^2,
\end{eqnarray}
this expression is obtained by computing the energy flux through a sphere containing the Q-Ball. When the real part of $k_\ep$ equals zero the fraction
becomes equal to one. The result in this range will be proportional to $\om^3_0$  \cite{Qballscohen,Qballsnew}. 
\subsubsection{Results of numerical integration}
We can now give the evaporation rate of a Q ball into massless fermions in function of its internal frequency.
\begin{figure}
\begin{center}
\includegraphics{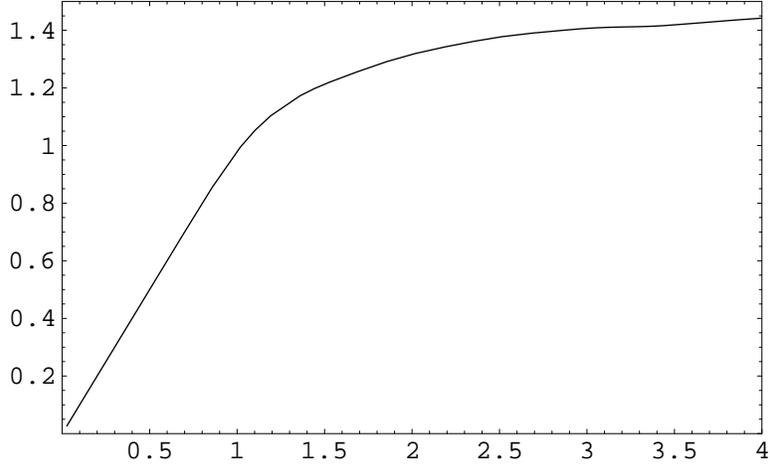}
\caption{Particle production rate $\frac{2\pi dN}{Mdt}$ as a function of $\frac{\om_0}{2M}$ in the limit of very big $Ml$ parameter.}
\label{figcurrent}
\end{center}
\end{figure}
In the first figure we can observe a limit in the evaporation rate. The absolute upper bound can be computed using
\begin{eqnarray}
\frac{dN}{dt}\leq\int_{-\frac{\om_0}{2}}^{+-\frac{\om_0}{2}}d\ep=\om_0, \nonumber
\end{eqnarray}
this absolute upper bound will be used to normalise the evaporation rate.
\begin{figure}
\begin{center}
\includegraphics{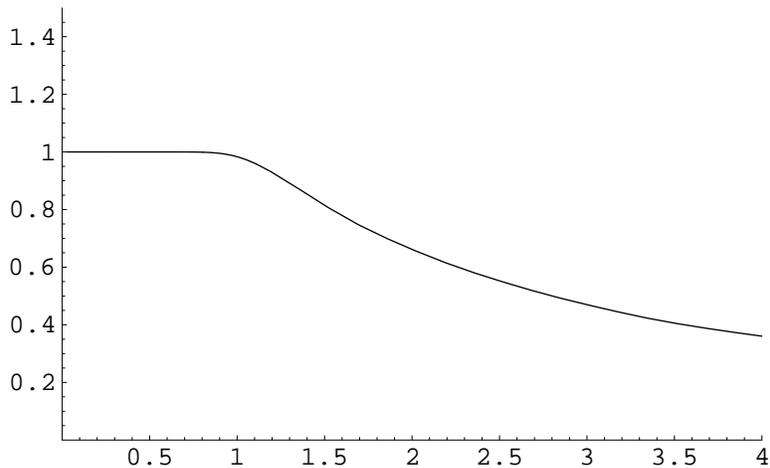}
\caption{Normalised particle production rate $\frac{2\pi dN}{Mdt}\frac{1}{\mbox{upper bound}}$ as a function of 
$\frac{\om_0}{2M}$ in the limit of a very big $Ml$ parameter.}
\label{figcurrent1}
\end{center}
\end{figure}
\section{Production of massive fermions}
The evaporation of a Q-ball into massive fermions is more complicated than the previous case. We can quite easily obtain the Heisenberg
field operator but solving the evaporation condition is a difficult task, even with only one space dimension. So the method we are going to use
is the same $S$-matrix based method used in \cite{Qballscohen,Qballsnew}. This picture will need as starting point the expression of the solution
as a superposition of wave packets. It is done by expressing the motion equations in matrix form and then expanding the solutions
over the eigenfunctions. This gives the separation into left and right movers. Choosing which wave is the incident one, we can
write the solution as
\begin{eqnarray}
\Psi_L=[B_1\Exp{i\bar{p}_1x}u_{\bar{p_1}}+B_2\Exp{i\bar{p}_2x}u_{\bar{p_2}}
+r_1\Exp{-i\bar{p}_1x}u_{-\bar{p_1}}+r_2\Exp{-i\bar{p}_2x}u_{-\bar{p_4}}], \\
\Psi_R=[t_1\Exp{i\bar{p}_1x}u_{\bar{p_1}}+t_2\Exp{i\bar{p}_2x}u_{\bar{p_2}}],
\end{eqnarray}
where the $u$'s and $\bar{p}$'s describe the solution away from the Q-ball and the $L$, $R$ subscripts stand for the left- or right-hand
side of the Q-ball. This solution has two incident waves associated with particles or anti-particles moving towards the Q-ball, giving two
solutions. The reflected and transmitted waves are associated with particles moving away from the Q-ball.
The same construction is done on the other side of the Q-ball, to  give four solutions. Finally we obtain the total solution
as a superposition of  these four solutions with the expansion coefficients becoming operators. This canonical quantisation does not 
introduce any big problem and can be done in a straightforward way. The next step will be to consider that in the far past only the
incoming wave survives, giving us a relation between the operators in the far past and in the far future (where only outgoing waves survives).
The last step we shall do is compute the number operator. 

The only difficult task is the computation of reflection and transmission amplitudes appearing in the solutions. Will shall provide
two methods for calculating these amplitudes. One method will consist in calculating all the scalar products of the motion eigenvectors,
while the other one will consist in the diagonalisation of the motion matrices. The results are fully consistent, and the two methods serve to
illustrate a variety of physical insights. The main objective we shall reach is the computation of the new value of the upper bound on evaporation 
rate. This bound will now depend on the produced fermion mass and not only on the internal Q-ball frequency.

The first three subsections describe the solutions in terms of eigenvectors of $4\times4$ matrices.
In the three next subsections we compute the transmission and reflection amplitudes for massive waves on the Q-Ball's surface. 
In the last subsections we compute the evaporation rate and simplify the problem by diagonalisation of the motion matrices.
\subsection{Preliminaries}
Using the same Lagrangian as for the massless case and adding a Dirac mass coupling for massive fermions, gives the fermionic
Lagrangian : 
\begin{eqnarray}
\LL=\bar{\psi}i\g^\mu\pam\psi+g(\bar{\psi}^C\psi\phi+h.c.)+M_D(x)\bar{\psi}\psi.
\end{eqnarray}
The equations of motion using two component $\psi$-field, 
in $1\oplus1$-dimensions are
\begin{eqnarray}
\ba{c}(i\pa_t+i\pa_z)\psi_1-M\Exp{-i\frac{\om_0}{2}t}\psi_2^\star+M_D\psi_2=0, \\
(i\pa_t-i\pa_z)\psi_2+M\Exp{-i\frac{\om_0}{2}t}\psi_1^\star+M_D\psi_1=0. \ea \label{mass1}
\end{eqnarray}
These equations can be solved if we use fields with four degrees of freedom. Solving this system will give us the
solution inside the Q-Ball, which is the first step. To solve these equations of motion we use 
the following ansatz
\begin{eqnarray}
\ba{c}\psi_1=f_1(z)\Exp{i(\ep-\frac{\om_0}{2})t}+f_2(z)\Exp{-i(\ep+\frac{\om_0}{2})t}, \\
\psi_2=g_1(z)\Exp{i(\ep-\frac{\om_0}{2})t}+g_2(z)\Exp{-i(\ep+\frac{\om_0}{2})t}. \ea \label{mass2}
\end{eqnarray}
Due to the separation of time components this ansatz will lead to four equations.
These equations can easily be modified to reduce the numbers of parameters : we divide all equations by $M\neq0$.
We shall now re-write these equations,
taking
\begin{eqnarray}
f_1(z)=A\Exp{ipz}, f_2^\star(z)=B\Exp{ipz}, g_1(z)=C\Exp{ipz} g_2^\star(z)=D\Exp{ipz}. 
\end{eqnarray}
After some
re-arrangement of the equations we obtain :
\begin{eqnarray}
\ba{c}-\ep_-f_1-g_2^\star+M_Dg_1=pf_1, \\
\ep_-g_1-f_2^\star-M_Df_1=pg_1, \\
-\ep_+f_2^\star+g_1-M_Dg_2^\star=pf_2^\star, \\
\ep_+g_2^\star+f_1+M_Df_2^\star=pg_2^\star,\ea
\end{eqnarray}
where $\ep_-=\ep-\frac{\om_0}{2}$ and $\ep_+=\ep+\frac{\om_0}{2}$.
This arrangement has the advantage that we can now write the $\psi$-field in terms of four component spinors as :
\begin{eqnarray}
\Psi=\lp\ba{c}\lp\ba{c}f_1\\g_1\ea\rp\\
\lp\ba{c}f_2^\star\\g_2^\star\ea\rp\ea\rp.
\end{eqnarray}
The idea of using four-component spinors is that now the fermion field contains both energy components,
just like the solution used in the previous chapter. The other advantage is that this rearrangement leads to the standard
four component spinor solution.
The equations of motion become in matrix form,
\begin{eqnarray}
\lp\ba{cccc} -\ep_- & M_D & 0 & -1 \\ -M_D & \ep_-& -1 & 0 \\
0& 1 & -\ep_+& -M_D \\ 1 & 0 & M_D & \ep_+\ea\rp\lp\ba{c} A \\B \\ C \\ D \ea\rp= 
Mp\lp\ba{c} A \\ B \\ C \\ D \ea\rp.\label{qballmatrix1}
\end{eqnarray}
All the parameters are normalised by $M$.
The fact that we have the $M$ factor one the right hand side will allows us to simplify the space components and replace
$l$ by $Ml$. All the parameters we are left with now are all dimensionless, we should read the matrix elements to be
all divided by $M$ and thus dimensionless. These satisfy
\begin{eqnarray}
\tau M \tau=M^T \label{sym},
\end{eqnarray}
with $\tau=DiagonalMatrix[1,-1,1,-1]$. This symmetry will be used to perform the normalisation of eigenvectors.
\subsection{Solution inside the Q-Ball}\label{solqball}
Using the dimensionless parameters and  the eigenvectors of the motion matrix we
can write the time independent solution inside the Q-Ball in the form :
\begin{eqnarray}
\Psi_Q=\sum_{j=1}^{4}C_jv_j\Exp{ip_jz},
\end{eqnarray}
The $v$'s are the eigenvectors of the motion matrix while the $p$'s are its eigenvalues. We shall not compute here the exact form of these
eigenvectors, but we can prove that the eigenvalues are either purely imaginary or purely real.
For reasons that will become clear later on, the first two terms of this solution have positive
momentum while the two last have negative momentum. This arrangement does not modify the shape
or any properties of the solution, but will greatly simplify the rest of the work. Inside the Q-ball the time dependent solution is :
\begin{eqnarray}
\Psi=\sum_{j=1}^{4}C_j\Exp{i(\ep-\om)t}v_{p_j}^{up}\Exp{i\bar{p}_jz}+C^\star_j\Exp{-i(\ep+\om)t}(v_{p_j}^{down}\Exp{i\bar{p}_jz})^\star,
\end{eqnarray}
where the $up$ superscript stands for the first two components of the eigenvectors, while the $down$ one indicates we
take the two last components.
\subsection{Solution without Q-Ball background.}
Outside the Q-Ball the solution is given by the eigenvalues and eigenvectors of the following 
matrix :
\begin{eqnarray}
\lp\ba{cccc} -\ep_- & M_D & 0 & 0 \\ -M_D & \ep_-& 0 & 0 \\
0& 0 & -\ep_+& -M_D \\ 0 & 0 & M_D & \ep_+\ea\rp\lp\ba{c} A \\B \\ C \\ D \ea\rp= 
p\lp\ba{c} A \\ B \\ C \\ D \ea\rp.\label{freematrix}
\end{eqnarray}
All its parameters are normalised by $M$,
so they are all dimensionless. The eigenvalues are given by :
\begin{eqnarray}
\ba{c}\bar{p}_{1,3}=\pm\sqrt{\ep_-^2-M_D^2}\equiv\pm\bar{p}_1,  \\
\bar{p}_{2,3}=\pm\sqrt{\ep_+^2-M_D^2}\equiv\pm\bar{p}_2, \ea
\end{eqnarray}
Let us stress that all  parameters are dimensionless since we have to read them as being divided by $M$, the Majorana mass coupling.
The exact eigenvectors can be easily calculated but we do not need them.
Here  the $\bar{p}_{1,2}$ momentum can be complex or real. If we want some particle to propagate outside Q-ball we need both $\bar{p}_{1,2}$ 
to be real, it gives for $\ep$ 
\begin{eqnarray}
|\ep_-|\geq M_D, \nonumber
\end{eqnarray}
to solve this we must identify two cases. The first case is,
\begin{eqnarray}
\ep_-\geq0\Rightarrow\ep\geq\frac{\om_0}{2} \nonumber \\
\ep_-\geq M_D\Rightarrow\ep\geq M_D+\frac{\om_0}{2}>\frac{\om_0}{2}, \nonumber
\end{eqnarray}
the last inequality is verified if $\frac{\om_0}{2}\geq M_D$. The second case is,
\begin{eqnarray}
\ep_-\leq0\Rightarrow\ep\leq\frac{\om_0}{2} \nonumber \\
-\ep_-\geq M_D\Rightarrow\frac{\om_0}{2}\geq\frac{\om_0}{2}-M_D\geq\ep \nonumber,
\end{eqnarray}
once more the last inequality is valid when $\frac{\om_0}{2}\geq M_D$. 
A similar calculation for $|\ep_+|\geq M_D$ gives :
\begin{eqnarray}
\ep\geq M_D-\frac{\om_0}{2} \quad\mbox{and}\quad \ep\leq-M_D-\frac{\om_0}{2}.
\end{eqnarray}
\begin{figure}
\begin{center}
\begin{picture}(0,0)%
\includegraphics{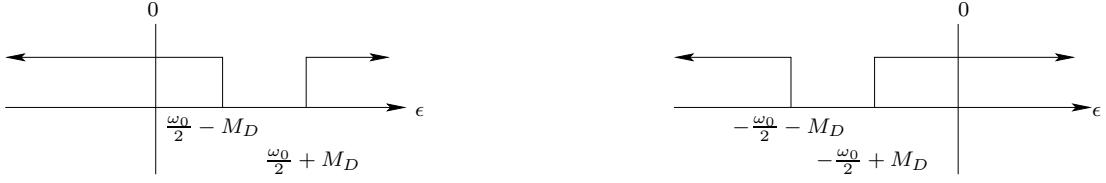}%
\end{picture}%
\setlength{\unitlength}{2763sp}%
\begingroup\makeatletter\ifx\SetFigFont\undefined%
\gdef\SetFigFont#1#2#3#4#5{%
  \reset@font\fontsize{#1}{#2pt}%
  \fontfamily{#3}\fontseries{#4}\fontshape{#5}%
  \selectfont}%
\fi\endgroup%
\begin{picture}(9774,1569)(889,-1423)
\put(4576,-886){\makebox(0,0)[lb]{\smash{{\SetFigFont{8}{9.6}{\rmdefault}{\mddefault}{\updefault}$\ep$}}}}
\put(8176,-1336){\makebox(0,0)[lb]{\smash{{\SetFigFont{8}{9.6}{\rmdefault}{\mddefault}{\updefault}$-\frac{\om_0}{2}+M_D$}}}}
\put(10651,-886){\makebox(0,0)[lb]{\smash{{\SetFigFont{8}{9.6}{\rmdefault}{\mddefault}{\updefault}$\ep$}}}}
\put(2176, 14){\makebox(0,0)[lb]{\smash{{\SetFigFont{8}{9.6}{\rmdefault}{\mddefault}{\updefault}$0$}}}}
\put(9451, 14){\makebox(0,0)[lb]{\smash{{\SetFigFont{8}{9.6}{\rmdefault}{\mddefault}{\updefault}$0$}}}}
\put(2326,-1036){\makebox(0,0)[lb]{\smash{{\SetFigFont{8}{9.6}{\rmdefault}{\mddefault}{\updefault}$\frac{\om_0}{2}-M_D$}}}}
\put(3226,-1336){\makebox(0,0)[lb]{\smash{{\SetFigFont{8}{9.6}{\rmdefault}{\mddefault}{\updefault}$\frac{\om_0}{2}+M_D$}}}}
\put(7426,-1036){\makebox(0,0)[lb]{\smash{{\SetFigFont{8}{9.6}{\rmdefault}{\mddefault}{\updefault}$-\frac{\om_0}{2}-M_D$}}}}
\end{picture}%
\end{center}
\caption{Sketch of the possible ranges for $\ep$}
\end{figure}
The only way to avoid the gaps and have the two waves (both $p_1$ and $p_2$) is for $\ep$ to be
in the range :
\begin{eqnarray}
\ep\in[M_D-\frac{\om_0}{2},\frac{\om_0}{2}-M_D].
\end{eqnarray}
This range will mix both the particles and anti-particles and thus lead to the non-trivial Bogolubov transformation.
We also have :
\begin{eqnarray}
\frac{\om_0}{2}\geq M_D.
\end{eqnarray}
This range is the equivalent as the range defined for the massless case.
Following the same construction as in the previous section,
the static solution outside the Q-Ball can also be written in the form
\begin{eqnarray}
\Psi_0=\sum_{j=1}^{4}A_ju_j\Exp{i\bar{p}_jz}, \\
\Psi_0=\sum_{j=1}^{4}B_ju_j\Exp{i\bar{p}_jz},
\end{eqnarray}
this time the $B_j$ coefficients are on the left-hand side of Q-Ball while the $A_j$ are
on the right hand side. The time dependent solution is once more given by,
\begin{eqnarray}
\Psi=\sum_{j=1}^{4}(A_j,B_j)\Exp{i(\ep-\om)t}u_{p_j}^{up}\Exp{i\bar{p}_jz}+(A_j^\star,B_j^\star)\Exp{-i(\ep+\om)t}(u_{p_j}^{down}\Exp{i\bar{p}_jz})^\star.
\end{eqnarray}
These considerations allow us to find the particle production energy range where particles can propagate outside the Q-ball.
The first part of our calculation is over, the next task  is to compute the reflection and transmission amplitudes. It is in fact solving
the matching equations in matrix form. The solution will give a relation from the far left to the far right of the Q-ball and the centre part (the Q-ball
itself) will not appear directly. 
\subsection{Construction of scattering matrix}
We want to construct the matrix linking the solution at $z=-\infty$ to the solution at $z=+\infty$.
We are searching for the matrix:
\begin{eqnarray}
\lp\ba{c}B_1 \\ B_2 \\ B_3 \\ B_4\ea\rp=
{\mathcal{V}}\lp\ba{c}A_1 \\ A_2 \\ A_3 \\ A_4\ea\rp. \label{diffusionV}
\end{eqnarray}
As a reminder the $B$'s are on the left while the $A$'s are the right hand side of the Q-Ball. This construction will not contain directly any
inside parameters so the $A$ and $B$ can be considered as free parameters while the reflection and transmission amplitudes will be contained in the
${\mathcal{V}}$ matrix. To compute the matrix elements we need to solve the matching equations. 
\subsection{Matching in space}
We first start by matching the solutions at $z=-l$ we have:
\begin{eqnarray}
B_1u_{\bar{p}_1}\Exp{-i\bar{p}_1l}+B_2u_{\bar{p}_2}\Exp{-i\bar{p}_2l}
+B_3u_{\bar{p}_3}\Exp{-i\bar{p}_3l}+B_4u_{\bar{p}_4}\Exp{-i\bar{p}_4l}= \nonumber \\
C_1v_{p_1}\Exp{-ip_1l}+C_2v_{p_2}\Exp{-ip_2l}+C_3v_{p_3}\Exp{-ip_3l}+C_4v_{p_4}\Exp{-ip_4l}\label{match1}
\end{eqnarray}
We redefine  the $B_i$ and the $C_i$ in the way
\begin{eqnarray}
\tilde{B}_i=B_i\Exp{-i\bar{p}_il}, \quad \tilde{C}_i=\frac{C_i}{\sqrt{N_i}}.
\end{eqnarray}
Multiplying equation (\ref{match1}) by $\tilde{u}_i^T\tau$, orthogonality property coming from the symmetry of the motion
matrices :
\begin{eqnarray}
\tilde{B}_iu^T_{p_i}\tau u_{p_i}=\sum_{j=1}^{4}u_i^T\tau v_j\Exp{-ip_jl}\tilde{C}_j, \label{match1}
\end{eqnarray}
Doing the same for all the $B$'s and writing down all the relations in matrix form we obtain :
\begin{eqnarray}
U\lp\ba{c}B_1\\B_2\\B_3\\B_4\ea\rp=
S{\mathcal E}\lp\ba{c}C_1\\C_2\\C_3\\C_4\ea\rp,
\end{eqnarray}
with,
\begin{eqnarray}
S=\lp\ba{cccc}u_1^T\tau v_1&u_1^T\tau v_2&u_1^T\tau v_3&u_1^T\tau v_4 \\
u_2^T\tau v_1&u_2^T\tau v_2&u_2^T\tau v_3&u_2^T\tau v_4 \\
u_3^T\tau v_1&u_3^T\tau v_2&u_3^T\tau v_3&u_3^T\tau v_4 \\
u_4^T\tau v_1&u_4^T\tau v_2&u_4^T\tau v_3&u_4^T\tau v_4 \ea\rp,
\end{eqnarray}
\begin{eqnarray}
{\mathcal E}(l)=\lp\ba{cccc}\Exp{-ip_1l}&0&0&0\\0&\Exp{-ip_2l}&0&0\\0&0&\Exp{-ip_3l}&0\\0&0&0&\Exp{-ip_4l}\ea\rp,
\label{prop}
\end{eqnarray}
and
\begin{eqnarray}
U=\lp\ba{cccc}u_{\bar{p}_1}^T\tau u_{\bar{p}_1}&0&0&0\\0&u_{\bar{p}_2}^T\tau u_{\bar{p}_2}&0&0\\0&0&u_{\bar{p}_3}^T\tau u_{\bar{p}_3}&0
\\0&0&0&u_{\bar{p}_4}^T\tau u_{\bar{p}_4}\ea\rp,
\label{norm}
\end{eqnarray}
We can then write for the expression we obtain at $z=-l$
\begin{eqnarray}
\lp\ba{c} B_1 \\ B_2 \\ B_3 \\ B_4 \ea\rp=U^{-1}S{\mathcal E}\lp\ba{c} C_1 \\ C_2 \\ C_3 \\ C_4 \ea\rp.
\end{eqnarray}
At $z=+l$ we have using the same definitions as for before :
\begin{eqnarray}
U\lp\ba{c} A_1 \\ A_2 \\ A_3 \\ A_4 \ea\rp=S{\mathcal E}(-l)\lp\ba{c} C_1 \\ C_2 \\ C_3 \\ C_4 \ea\rp,
\end{eqnarray}
Mixing up these two relations we obtain for the total transformation matrix ${\mathcal V}$ the following relation :
\begin{eqnarray}
\lp\ba{c} B_1 \\ B_2 \\ B_3 \\ B_4 \ea\rp=U^{-1}S{\mathcal E}{{\mathcal E}'}^{-1}S^{-1}U\lp\ba{c} A_1 \\ A_2 \\ A_3 \\ A_4 \ea\rp.
\end{eqnarray}
Using the definition
\begin{eqnarray}
{\mathcal E(l)}{{\mathcal E}(-l)}^{-1}\equiv E,
\end{eqnarray}
we can easily show that :
\begin{eqnarray}
[E,\tau]_-=0.
\end{eqnarray}
As we shall find out later on the last form is a transformation that allows us to diagonalise the motion
matrix inside the Q-ball.
Using this matrix we shall construct all the reflection and diffusion coefficients for all the
waves moving inside and outside of the Q-Ball. Before we continue we need to remember that $p_3=-p_1$
and $p_4=-p_2$ for both sets of $p$'s (barred ones and no bar ones). We see here that the choice for normalisation of 
eigenvectors will just act on the $U$ matrix that can be either the identity or the $\tau$ matrix or any other choice we can make.
Finally the diffusion matrix $V$ we where looking for is given by :
\begin{eqnarray}
V=U^{-1}SES^{-1}U.
\end{eqnarray}
In fact nothing we shall do depends on these matrices but we have shown the full procedure for completeness.
\subsection{Construction of reflection and transmission amplitudes}
We shall first construct the reflection and transmission amplitudes from the {\em left side to the right hand 
side} of the 
Q-Ball. To do so we shall use the definition of the
${\mathcal V}$ matrix given by equation \ref{diffusionV}. The first two coefficients are linked to right-moving waves while the last two
coefficients are linked to left-moving waves. This choice for arranging the waves was done for simplicity, we can
show that any other arrangement  leads to same results. In fact, we choose the simplest possible arrangement.
Due to the shape of the $u$ spinors and the $\Omega$ matrix in front the first and the third coefficients of the free solution have the same energy 
while the second and the fourth coefficients correspond to another energy wave. We shall identify these two
energy ranges to the type one particles (1) and type two particles (2). Using equation \ref{diffusionV}
and separating the matrix into four two by two blocs we can write :
\begin{eqnarray}
\lp\ba{c} \rightarrow \\ \rightarrow \\ r_1 \\ r_2 \ea\rp=\lp\ba{c|c} {\mathcal V}_{11} & {\mathcal V}_{12} \\
\hline  {\mathcal V}_{21} & {\mathcal V}_{22} \ea\rp\lp\ba{c} t_1 \\ t_2 \\ 0 \\ 0\ea\rp,
\end{eqnarray}
where the two arrows stand for the incoming waves, the first two coefficients will be replaced by one.
Using the bloc separation of the matrix we find :
\begin{eqnarray}
\lp\ba{c} \rightarrow \\ \rightarrow \ea\rp={\mathcal V}_{11}\lp\ba{c} t_1 \\ t_2 \ea\rp, \\
\lp\ba{c} r_1 \\ r_2 \ea\rp={\mathcal V}_{21}\lp\ba{c} t_1 \\ t_2 \ea\rp,
\end{eqnarray}
leading to
\begin{eqnarray}
\lp\ba{c} t_1 \\ t_2 \ea\rp=\underbrace{{\mathcal V}_{11}^{-1}}_{{\mathcal T}}\lp\ba{c} \rightarrow \\ \rightarrow \ea\rp, \\
\lp\ba{c} r_1 \\ r_2 \ea\rp=\underbrace{{\mathcal V}_{21}{\mathcal V}_{11}^{-1}}_{{\mathcal R}}
\lp\ba{c} \rightarrow \\ \rightarrow \ea\rp.
\end{eqnarray} 
The ${\mathcal R}$ and ${\mathcal T}$ matrices will give the reflection and transmission amplitudes
when they are applied on the incoming wave coefficients. These two matrices are two by two the first line
corresponding to transmission or reflection of particles with  two different incoming waves, while the
second line gives the coefficients for anti-particles.
We shall construct the transmission and reflection coefficients from the {\em right to the left hand side} of Q-Ball.
Using the same method as before we have this time :
\begin{eqnarray}
\lp\ba{c} 0 \\ 0 \\ \tilde{t}_1 \\ \tilde{t}_2 \ea\rp=\lp\ba{c|c} {\mathcal V}_{11} & {\mathcal V}_{12} \\
\hline  {\mathcal V}_{21} & {\mathcal V}_{22} \ea\rp\lp\ba{c} \tilde{r}_1 \\ \tilde{r}_2 \\ \leftarrow \\
\leftarrow \ea\rp,
\end{eqnarray}
leading this time to
\begin{eqnarray}
\lp\ba{c} 0 \\ 0 \ea\rp={\mathcal V}_{11}\lp\ba{c} \tilde{r}_1 \\ \tilde{r}_2\ea\rp +{\mathcal V}_{12}
\lp\ba{c} \leftarrow \\ \leftarrow \ea\rp,\\
\lp\ba{c} \tilde{t}_1 \\ \tilde{t}_2 \ea\rp={\mathcal V}_{21}\lp\ba{c} \tilde{r}_p \\ \tilde{r}_{a-p}\ea\rp +{\mathcal V}_{22}
\lp\ba{c} \leftarrow \\ \leftarrow \ea\rp.
\end{eqnarray}
Solving these two equations gives the reflection and transmission matrices for incoming particles from the left,
they are :
\begin{eqnarray}
\tilde{{\mathcal R}}=-{\mathcal V}_{12}{\mathcal V}_{11}^{-1}, \\
\tilde{{\mathcal T}}={\mathcal V}_{22}-{\mathcal V}_{21}{\mathcal V}_{11}^{-1}{\mathcal V}_{12}.
\end{eqnarray}
Now that all the coefficients are known we can construct and quantise the solution.
\subsection{Construction of standard solution}
\begin{figure}
\begin{center}
\begin{picture}(0,0)%
\includegraphics{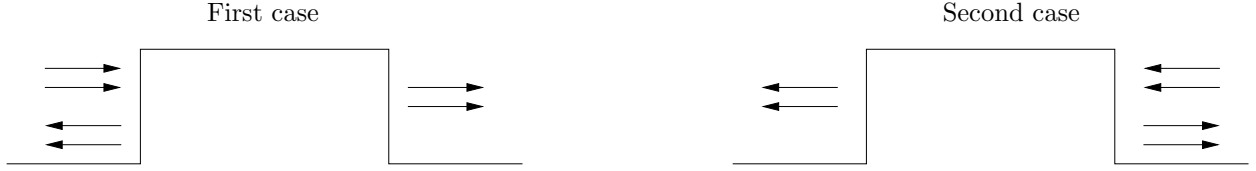}%
\end{picture}%
\setlength{\unitlength}{3158sp}%
\begingroup\makeatletter\ifx\SetFigFont\undefined%
\gdef\SetFigFont#1#2#3#4#5{%
  \reset@font\fontsize{#1}{#2pt}%
  \fontfamily{#3}\fontseries{#4}\fontshape{#5}%
  \selectfont}%
\fi\endgroup%
\begin{picture}(9774,1272)(589,-1423)
\put(2176,-286){\makebox(0,0)[lb]{\smash{{\SetFigFont{10}{12.0}{\rmdefault}{\mddefault}{\updefault}First case}}}}
\put(7951,-286){\makebox(0,0)[lb]{\smash{{\SetFigFont{10}{12.0}{\rmdefault}{\mddefault}{\updefault}Second case}}}}
\end{picture}%
\end{center}
\caption{Sketch of both cases used to build the solution : we have each time two incident particles, two reflected and two
transmitted. It is an effect of massive particles because we can not identify any more the particles with the anti-particles.}
\label{figmassive1}
\end{figure}

Using the transmission and reflection coefficients we can identify two different cases the first case is
when incident particles are on the left hand side of the Q-Ball, while the other case stands for incident particles 
coming from the right hand side (see fig. \ref{figmassive1}).
We shall treat separately the solution on the left and the solution on the right, the matching coefficients are those found
in the previous section, they link the expansion coefficients on the right to those on the left.
Writing down these two possibilities we have :
\begin{eqnarray}
\Psi_L=[B_1\Exp{i\bar{p}_1x}u_{\bar{p_1}}+B_2\Exp{i\bar{p}_2x}u_{\bar{p_2}}
+r_1\Exp{-i\bar{p}_1x}u_{-\bar{p_1}}+r_2\Exp{-i\bar{p}_2x}u_{-p_4}], \\
\Psi_R=[t_1\Exp{i\bar{p}_1x}u_{\bar{p_1}}+t_2\Exp{i\bar{p}_2x}u_{\bar{p_2}}],
\end{eqnarray}
for the first case, the two incident particles coming from the left hand side of the Q-Ball and 
\begin{eqnarray}
\Psi_L=[\tilde{t}_1\Exp{-i\bar{p}_1x}u_{-\bar{p_1}}+\tilde{t}_2\Exp{-i\bar{p}_2x}u_{-\bar{p_2}}],\\
\Psi_R=[\tilde{r}_1\Exp{i\bar{p}_1x}u_{\bar{p_1}}+\tilde{r}_2\Exp{i\bar{p}_2x}u_{\bar{p_2}}
+A_1\Exp{-i\bar{p}_1x}u_{-\bar{p_1}}+A_2\Exp{-i\bar{p}_2x}u_{-\bar{p_2}}],
\end{eqnarray}
for the second case. In both of these definitions we have :
\begin{eqnarray}
\lp\ba{c}r_1 \\ r_2 \ea\rp={\mathcal R}\lp\ba{c} B_1 \\ B_2 \ea\rp \quad 
\lp\ba{c}t_1 \\ t_2 \ea\rp={\mathcal T}\lp\ba{c} B_1 \\ B_2 \ea\rp, \label{coef1}\\
\lp\ba{c}\tilde{r}_1 \\ \tilde{r}_2 \ea\rp=\tilde{{\mathcal R}}\lp\ba{c} A_1 \\ A_2 \ea\rp \quad 
\lp\ba{c}\tilde{t}_1 \\ \tilde{t}_2 \ea\rp=\tilde{{\mathcal T}}\lp\ba{c} A_1 \\ A_2 \ea\rp\label{coef2}.
\end{eqnarray}
To clearly understand the construction, the $B$'s are the incident amplitudes from the left while the $A$'s
are the amplitudes from the right. The $1$ and $2$ subscript indicate the type of particle we are dealing with,
we have two different exponentials in $\Omega(t)$. We then need to take the complex conjugate of the terms
corresponding to the two last components of spinors. To continue building the solution we still need to
separate each of these two cases in two, considering only one type of incident particle at the time.
This construction leads to the four following pieces, that will be identified to the four degrees of freedom
that our solution has :
\begin{eqnarray}
\ba{c}
\Psi_L=[\Exp{i\bar{p}_1x}u_{\bar{p_1}}+r_{11}\Exp{-i\bar{p}_1x}u_{-\bar{p_1}}]+[r_{12}\Exp{-i\bar{p}_2 x}u_{-\bar{p_2}}], \\
\Psi_R=[t_{11}\Exp{i\bar{p}_1x}u_{\bar{p_1}}]+[t_{12}\Exp{i\bar{p}_2 x}u_{\bar{p_2}}], \ea
\end{eqnarray}
for the one incident type one particle from the left and
\begin{eqnarray}
\ba{c}\Psi_L=[r_{21}\Exp{-i\bar{p}_1x}u_{-\bar{p_2}}]+[\Exp{i\bar{p}_2 x}u_{\bar{p_2}}+r_{22}\Exp{-i\bar{p}_2 x}u_{-\bar{p_2}}], \\
\Psi_R=[t_{21}\Exp{i\bar{p}_1x}u_{\bar{p_1}}]+[t_{22}\Exp{-i\bar{p}_2 x}u_{\bar{p_2}}], \ea
\end{eqnarray}
for an incident type two particle. The coefficients are given by :
\begin{eqnarray}
\lp\ba{c}r_{11} \\ r_{12} \ea\rp={\mathcal R}\lp\ba{c} 1 \\ 0 \ea\rp \quad 
\lp\ba{c}t_{11} \\ t_{12} \ea\rp={\mathcal T}\lp\ba{c} 1 \\ 0 \ea\rp, \label{coef3}\\
\lp\ba{c}r_{21} \\ r_{22} \ea\rp={\mathcal R}\lp\ba{c} 0 \\ 1 \ea\rp \quad 
\lp\ba{c}t_{21} \\ t_{22} \ea\rp={\mathcal T}\lp\ba{c} 0 \\ 1 \ea\rp\label{coef4}.
\end{eqnarray}
The two other pieces for particles incident from the right we have :
\begin{eqnarray}
\ba{c}\Psi_L=[\tilde{t}_{11}\Exp{-i\bar{p}_1x}u_{-\bar{p_1}}]+[\tilde{t}_{12}\Exp{-i\bar{p}_2 x}u_{-\bar{p_2}}], \\
\Psi_R=[\tilde{r}_{11}\Exp{i\bar{p}_1x}u_{\bar{p_1}}+\Exp{-i\bar{p}_1x}u_3]+[\tilde{r}_{12}\Exp{-i\bar{p}_2 x}u_2],\ea
\end{eqnarray}
for one incident type one particle from the right and
\begin{eqnarray}
\ba{c}\Psi_L=[\tilde{t}_{21}\Exp{-i\bar{p}_1x}u_{-\bar{p_1}}]+[\tilde{t}_{22}\Exp{-i\bar{p}_2 x}u_{-\bar{p_2}}], \\
\Psi_R=[\tilde{r}_{21}\Exp{i\bar{p}_1x}u_{\bar{p_1}}]+[\tilde{r}_{22}\Exp{-i\bar{p}_2 x}u_{\bar{p_2}}+\Exp{-i\bar{p}_2 x}u_{-\bar{p_2}}],\ea
\end{eqnarray}
for an incident type two particle and finally the coefficients are given by :
\begin{eqnarray}
\lp\ba{c}\tilde{r}_{11} \\ \tilde{r}_{12} \ea\rp=\tilde{{\mathcal R}}\lp\ba{c} 1 \\ 0 \ea\rp \quad 
\lp\ba{c}\tilde{t}_{11} \\ \tilde{t}_{12} \ea\rp=\tilde{{\mathcal T}}\lp\ba{c} 1 \\ 0 \ea\rp, \\
\lp\ba{c}\tilde{r}_{21} \\ \tilde{r}_{22} \ea\rp=\tilde{{\mathcal R}}\lp\ba{c} 0 \\ 1 \ea\rp \quad 
\lp\ba{c}\tilde{t}_{21} \\ \tilde{t}_{22} \ea\rp=\tilde{{\mathcal T}}\lp\ba{c} 0 \\ 1 \ea\rp.
\end{eqnarray}
If we want to easily remember the coefficients there is an easy trick. For the coefficients without the tilde
we read the subscript from the left to the right, $r_{21}$ is the coefficient for an incident type two particle 
being reflected as a type one particle. The same lecture stands also for the tilde coefficients representing incident particles from the right. 
\subsection{Quantisation and Bogolubov transformation}
Quantisation of solution is now easy, the total quantised solution will be a linear combination of all four
parts, with expansion coefficients becoming operators after the normalisation is made. 
The solution is given by :
\begin{eqnarray}
\Psi=\sum_{j=1}^{4}(A_j,B_j)\Exp{i(\ep-\frac{\om_0}{2})t}u_{p_j}^{up}\Exp{i\bar{p}_jz}+
(A_j^\star,B_j^\star)\Exp{-i(\ep+\frac{\om_0}{2})t}(u_{p_i}^{down}\Exp{i\bar{p}_iz})^\star,
\end{eqnarray}
in our case only the eigenvectors corresponding to the $\bar{p_1}$ eigenvalue have $up$ components, while only the eigenvectors
corresponding to $\bar{p_2}$ have down components.
\begin{eqnarray}
\Psi_L=\Exp{i(\ep-\frac{\om_0}{2})t}[\Exp{i\bar{p}_1x}u_{\bar{p_1}}^{up}+r_{11}\Exp{-i\bar{p}_1x}u_{-\bar{p_1}}^{up}]+
\Exp{i(\ep+\frac{\om_0}{2})t}[r_{12}\Exp{-i\bar{p}_2 x}u_{\bar{p_2}}],
\end{eqnarray}
incident particle is the wave containing $u_{\bar{p_1}}$, all this superposition must be represented using the same
operator so we are sure to have only four degrees of freedom. Quantisation will be done using energy, for the incident wave
we have :
\begin{eqnarray}
\ba{c}\Exp{i(\ep-\frac{\om_0}{2})t}u_1\rightarrow\ep-\frac{\om_0}{2}\geq M_D, \\
\rightarrow\ep\geq M_D+\frac{\om_0}{2},\ea
\end{eqnarray}
leading for this first wave to 
\begin{eqnarray}
\Psi_L=\int_{M_D+\frac{\om_0}{2}}^\infty d\ep\{\Exp{i(\ep-\frac{\om_0}{2})t}[\Exp{i\bar{p}_1x}u_1^{up}+r_{11}\Exp{-i\bar{p}_1x}u_3^{up}]+
\nonumber \\ \Exp{i(\ep+\frac{\om_0}{2})t}[r_{12}\Exp{-i\bar{p}_2 x}(u_{-\bar{p_2}}^{down})^\star]\}b^\dagger(\bar{p}_1),
\end{eqnarray}
after the conjugation of the term proportional to $\Exp{i(\ep+\frac{\om_0}{2})t}$ we have,
\begin{eqnarray}
\Psi_L=\int_{M_D+\frac{\om_0}{2}}^\infty d\ep\{\Exp{i(\ep-\frac{\om_0}{2})t}[\Exp{i\bar{p}_1x}u_{\bar{p_1}}^{up}+r_{11}\Exp{-i\bar{p}_1x}
u_{-\bar{p_1}}^{up}]
b^\dagger_{\bar{p}_1}+\nonumber \\
\Exp{-i(\ep+\frac{\om_0}{2})t}[r_{12}^\star\Exp{i\bar{p}_2^\star x}(u_{-\bar{p_2}}^{down})^\star]b_{\bar{p}_1}\}.
\end{eqnarray}
Applying the same method to all the terms we finally obtain for the total solution having four degrees of freedom : 
{\small
\begin{eqnarray}
\Psi_L&=&\int_{M_D+\frac{\om_0}{2}}^\infty d\ep
\{\Exp{i(\ep-\frac{\om_0}{2})t}[\Exp{i\bar{p}_1x}u_{\bar{p_1}}^{up}+r_{11}\Exp{-i\bar{p}_1x}u_{-\bar{p_1}}^{up}]b^\dagger_{\bar{p}_1}  
+\Exp{-i(\ep+\frac{\om_0}{2})t}[r_{12}^\star\Exp{i\bar{p}_2^\star x}(u_{-\bar{p_2}}^{down})^\star]b_{\bar{p}_1}\} \nonumber\\
&+&\int_{M_D-\frac{\om_0}{2}}^\infty d\ep\{\Exp{i(\ep-\frac{\om_0}{2})t}[r_{21}\Exp{-i\bar{p}_1x}u_{-\bar{p_1}}^{up}]a^\dagger_{\bar{p}_2}
+\Exp{-i(\ep+\frac{\om_0}{2})t}[\Exp{-i\bar{p}_2^\star x}u_{\bar{p_2}}^\star+r_{22}^\star\Exp{i\bar{p}_2^\star x}(u_{-\bar{p_2}}^{down})^\star]
a_{\bar{p}_2}\} 
\nonumber\\
&+&\int_{M_D+\frac{\om_0}{2}}^\infty d\ep\{\Exp{i(\ep-\frac{\om_0}{2})t}[\tilde{t}_{11}\Exp{-i\bar{p}_1x}u_{-\bar{p_1}}^{up}]b^\dagger_{-\bar{p}_1} 
+\Exp{-i(\ep+\frac{\om_0}{2})t}[\tilde{t}^\star_{12}\Exp{i\bar{p}_2^\star x}(u_{-\bar{p_2}}^{down})^\star]b_{-\bar{p}_1}\} \nonumber\\
&+&\int_{M_D-\frac{\om_0}{2}}^\infty d\ep\{\Exp{i(\ep-\frac{\om_0}{2})t}[\tilde{t}_{21}\Exp{-i\bar{p}_1x}u_{-\bar{p_1}}^{up}]a^\dagger_{-\bar{p}_2} 
+\Exp{-i(\ep+\frac{\om_0}{2})t}[\tilde{t}^\star_{22}\Exp{i\bar{p}_2^\star x}(u_{-\bar{p_2}}^{down})^\star]a_{-\bar{p}_2}\}.\nonumber
\end{eqnarray}
}
The upper integration bound is $\frac{\om_0}{2}-M_D$ so we are only left with the terms containing the $a$ operators,
\begin{eqnarray}
\Psi_L&=&\int_{M_D-\frac{\om_0}{2}}^{\frac{\om_0}{2}-M_D} d\ep\{\Exp{i(\ep-\frac{\om_0}{2})t}[r_{21}\Exp{-i\bar{p}_1x}u_{-\bar{p_1}}^{up}
]a^\dagger_{\bar{p}_2}
+\Exp{-i(\ep+\frac{\om_0}{2})t}[\Exp{-i\bar{p}_2^\star x}(u_{\bar{p_2}}^{down})^\star+r_{22}^\star\Exp{i\bar{p}_2^\star x}
(u_{-\bar{p_2}}^{down})^\star]a_{\bar{p}_2}\} \nonumber \\ \\
&+&\int_{M_D-\frac{\om_0}{2}}^{\frac{\om_0}{2}-M_D} d\ep\{\Exp{i(\ep-\frac{\om_0}{2})t}[\tilde{t}_{21}\Exp{-i\bar{p}_1x}
u_{-\bar{p_1}}^{up}]a^\dagger_{-\bar{p}_2} 
+\Exp{-i(\ep+\frac{\om_0}{2})t}[\tilde{t}^\star_{22}\Exp{i\bar{p}_2^\star x}(u_{-\bar{p_2}}^{down})^\star]a_{-\bar{p}_2}\}\nonumber.
\end{eqnarray}
One interesting result can be found here is that like in the massless case if we change the sign of $\frac{\om_0}{2}$ 
we change the particle type since we change the operator type. For the moment the $a$ coefficients are only expansion coefficients since we
have not quantised the wave yet. The solution on the right hand side of the Q-Ball is given by :
\begin{eqnarray}
\Psi_{R}&=&\int_{M_D-\frac{\om_0}{2}}^{\frac{\om_0}{2}-M_D} d\ep\{[t_{12}\Exp{i\bar{p}_1x}u_{\bar{p_1}}]a^\dagger_{\bar{p}_2}+
[t_{22}^\star\Exp{-i\bar{p}_2^\star x}u_{\bar{p_2}}]a_{\bar{p}_2}\} \nonumber \\ \\
&+&\int_{M_D-\frac{\om_0}{2}}^{\frac{\om_0}{2}-M_D} d\ep\{[\tilde{r}_{12}\Exp{i\bar{p}_1 x}u_{\bar{p_1}}]a^\dagger_{-\bar{p}_2}+
[\tilde{r}_{22}^\star\Exp{-i\bar{p}_1 x}u_{\bar{p_2}}+\Exp{i\bar{p}_1 x}u_{-\bar{p_2}}]a{-\bar{p}_2} \}\nonumber
\end{eqnarray}
At $t=+\infty$ only the terms without any incident wave will survive we have then
\begin{eqnarray}
r_{22}^\star a(\bar{p}_2)+\tilde{t}_{22}^\star a(-\bar{p}_2)+r_{21}a^\dagger(\bar{p}_2)+\tilde{t}_{12}a^\dagger(-\bar{p}_2)
=a_{out}(\bar{p}_2),
\end{eqnarray}
this is the Bogolubov transformation we where looking for. If we want the $a_{out}$ coefficient to be a operator we need to check that it satisfies
the same anti-commutation relations as $a_{in}$. We have :
\begin{eqnarray}
\{(a_{out}^\dagger)',a_{out}\}&=&\lp(r_{21}^\star)'a_{\bar{p}_2'}+(r_{22}^\star)'a^\dagger_{\bar{p}_2'}+
(\tilde{t}_{21}^\star)'a_{-\bar{p}_2'}+(\tilde{t}_{22}^\star)'a^\dagger_{-\bar{p}_2'}\rp \nonumber \\
&\times&\lp(r_{12})a^\dagger_{\bar{p}_2}+(r_{22})a_{\bar{p}_2}+(t_{21})a^\dagger_{-\bar{p}_2}+
(t_{22})a_{-\bar{p}_2}\rp \nonumber \\
&=&\lp(r_{21}^\star)'r_{21}+(r_{22}^\star)'r_{22}+(\tilde{t}_{21}^\star)'\tilde{t}_{21}+(\tilde{t}_{22}^\star)'\tilde{t}_{22}\rp
\{(a_{in}^\dagger)',a_{in}\}. \nonumber
\end{eqnarray}
This relation  can also be obtained if we set that the incident current is equal to the outgoing one, or even with the normalisation
of wave packets. At this stage it can be important to use some normalised eigenvectors. As will shall show later on it is always the case
if we diagonalise the matrix outside the Q-Ball. The number of created particles is now given by
\begin{eqnarray}
_{in}<0|a^\dagger_{out}a_{out}|0>_{in}=\left(\frac{|r_{21}|^2+|\tilde{t}_{21}|^2}{|r_{21}|^2+|r_{22}|^2+|\tilde{t}_{21}|^2+|\tilde{t}_{22}|^2}
\right)\delta(\ep-\ep').
\end{eqnarray}
We need to smooth out this result, to do so we shall use the same argument as \cite{Qballscohen} to finally obtain
\begin{eqnarray}
\frac{dN}{dt}=\frac{1}{2\pi}\int_{M_D-\frac{\om_0}{2}}^{\frac{\om_0}{2}-M_D}\left(\frac{|r_{21}|^2+|\tilde{t}_{21}|^2}
{|r_{21}|^2+|r_{22}|^2+|\tilde{t}_{21}|^2+|\tilde{t}_{22}|^2}\right)d\ep.
\end{eqnarray}
Since we are dealing with a Bogolubov transformation we have 
\begin{eqnarray}
\left(\frac{|r_{21}|^2+|\tilde{t}_{21}|^2}{|r_{21}|^2+|r_{22}|^2+|\tilde{t}_{21}|^2+|\tilde{t}_{22}|^2}\right)\leq1, \\
\frac{dN}{dt}\leq\om_0-2M_D,
\end{eqnarray}
In fact the best thing to do is to consider the solution in all space on the left and on the right instead of considering only one
side. To do so we just need to consider an incident particle on the left and build the solution without tilde factors. The rest of the procedure 
is the same we have,
\begin{eqnarray}
\Psi_{L+R}&=&\int_{M_D-\frac{\om_0}{2}}^{\frac{\om_0}{2}-M_D} d\ep\{\Exp{i(\ep-\frac{\om_0}{2})t}[r_{21}\Exp{-i\bar{p}_1x}u_{-\bar{p_1}}^{up}
]a^\dagger_{\bar{p}_2}
+\Exp{-i(\ep+\frac{\om_0}{2})t}[\Exp{-i\bar{p}_2^\star x}(u_{\bar{p_2}}^{down})^\star+r_{22}^\star\Exp{i\bar{p}_2^\star x}
(u_{-\bar{p_2}}^{down})^\star]a_{\bar{p}_2}\}, \nonumber \\ \\
&+&\int_{M_D-\frac{\om_0}{2}}^{\frac{\om_0}{2}-M_D} d\ep\{\Exp{i(\ep-\frac{\om_0}{2})t}[t_{21}\Exp{i\bar{p}_1x}
u_{\bar{p_1}}^{up}]a^\dagger_{-\bar{p}_2} 
+\Exp{-i(\ep+\frac{\om_0}{2})t}[t^\star_{22}\Exp{-i\bar{p}_2^\star x}(u_{\bar{p_2}}^{down})^\star]a_{-\bar{p}_2}\}\nonumber,
\end{eqnarray}
this time leading to
\begin{eqnarray}
\frac{dN}{dt}=\frac{1}{2\pi}\int_{M_D-\frac{\om_0}{2}}^{\frac{\om_0}{2}-M_D}\left(\frac{|r_{21}|^2+|t_{21}|^2}
{|r_{21}|^2+|r_{22}|^2+|t_{21}|^2+|t_{22}|^2}\right)d\ep,
\end{eqnarray}
after normalisation of operators.
These results seem to be correct because when the fermions become massless there is identification of both types of produced particles
so the total coefficient becomes equal to one as in the previous section and there is total reflection.
We could stop our calculations here since we've got the expression of particle production state and all the matrix elements are
known. But we can greatly simplify this problem so the expressions we obtain become simpler and more readable.
\subsection{Direct construction of $S$ Matrix}
Using the shape of the different matrices we deal with we think there is a simpler way to construct the diffusion matrix.
In fact all the matrices of motion equations can be diagonalised using simple transformations that preserve the 
symmetry of the problem. If the matrices can be diagonalised the eigenvectors will have automatic orthogonality and 
normalisation properties. This diagonalisation will be done using simple transformations depending on six parameters. The 
symmetry to conserve is given in (\ref{sym}).

Taking a look at the $M_0$ matrix defined in eq. (\ref{freematrix}) we can diagonalise it using the Lorentz boost-transformation :
\begin{eqnarray}
M_0'=\tau v_1^T\tau M_0 v_1,
\end{eqnarray}
with,
\begin{eqnarray}
v_1&=&\lp\ba{cccc}\cosh(x_1) & \sinh(x1) & 0 & 0 \\ \sinh(x_1) & \cosh(x_1) & 0 & 0 \\
0 & 0 & \cosh(x_2) & -\sinh(x_2) \\0 & 0 & -\sinh(x_2) & \cosh(x_2)\ea\rp, \\
v_1^T\tau v_1&=&\tau.
\end{eqnarray}
The last equation ensures us the fact that $\tau v^T \tau=v^{-1}$ and that the symmetry of the problem is
conserved.
Setting $x_1$ and $x_2$ being solutions of :
\begin{eqnarray}
\ba{c}\cosh(2x_1)=\frac{\ep_-\sinh(2x_1)}{M_D},\\
\cosh(2x_2)=\frac{\ep_+\sinh(2x_2)}{M_D},\ea\label{diag2}
\end{eqnarray}
we find for the $M_0'$ matrix the following diagonal form,
\begin{eqnarray}
M_0'=\lp\ba{cccc} k_1 & 0 & 0 & 0 \\ 0 & -k_1 & 0 & 0 \\ 0 & 0 & k_2 & 0 \\ 0 & 0 & 0 & -k_2 \ea\rp,
\end{eqnarray}
where
\begin{eqnarray}
\ba{c}k_1=\frac{(M_D^2-\ep_-^2)\sinh(x_1)}{M_D}, \\
k_2=\frac{(M_D^2-\ep_+^2)\sinh(x_2)}{M_D}.\ea
\end{eqnarray} 
All the parameters we find after this transformation are real, the $k$'s that we find represent the momentum of the particles in
this new base.
The same transformation applied on the $M_1$ matrix defined by eq. (\ref{qballmatrix1}) gives :
\begin{eqnarray}
M_1'=\lp\ba{cccc}k_1&0&\sinh(x)&-\cosh(x)\\0&-k_1&-\cosh(x)&\sinh(x)\\
\sinh(x)&\cosh(x)&k_2&0\\\cosh(x)&\sinh(x)&0&-k_2\ea\rp\label{matadiag},
\end{eqnarray}
with $x=x_1+x_2$. We can check that this new matrix has the same symmetry properties as $M_1$. This simple transformation 
allows us to eliminate the Dirac coupling of our equations but in presence
of the Q-Ball it is replaced by a double Majorana coupling. Since this transformation is made every where, it will not change any of the properties.
A solution of equations (\ref{diag2}) is easy to construct it is :
\begin{eqnarray}
\frac{\cosh(2x_1)}{\sinh(2x_1)}=\coth(2x_1)=\frac{\ep_-}{M_D} \nonumber \\
2x_1={\mathrm{argcoth}}(\frac{\ep_-}{M_D})
\end{eqnarray}
Once this transformation is made inside and outside of the Q-Ball the eigenvectors outside of the Q-Ball only contain one non zero component,
the important fact here is that the $M_0'$ matrix is now self adjoint so its eigenvectors have the standard
orthogonality properties without the $\tau$ matrix.  When we do the matching in space eq. (\ref{match1}) instead
of multiplying by $u_i^T\tau$ we multiply by $u_i^\dagger$ and the $S$ matrix is made of the components
of $M_1'$ eigenvectors, so the only thing we did was to diagonalise the transformed matrix.
To continue the Diagonalisation process we now transform $M_1'$
in the way :
\begin{eqnarray}
M_1''=\tau (s_1s_2)^T\tau M_1'(s_1s_2)
\end{eqnarray}
with,
\begin{eqnarray}
s_1&=&
\lp\ba{cccc}\cosh[y/2]&0&0&\sinh[y/2]\\0&\cosh[y/2]&-\sinh[y/2]&0\\0&-\sinh[y/2]&\cosh[y/2]&0\\
\sinh[y/2]&0&0&\cosh[y/2]\ea\rp, \\
s_2&=&
\lp\ba{cccc}\cos[z/2]&0&\sin[z/2]&0\\0&\cos[z/2]&0&-\sin[z/2]\\-\sin[z/2]&0&\cos[z/2]&0\\
0&\sin[z/2]&0&\cos[z/2]\ea\rp. 
\end{eqnarray}
As before we have :
\begin{eqnarray}
(s_1s_2)^T\tau(s_1s_2)=\tau,
\end{eqnarray}
to preserve the symmetry of the problem. 
This set of transformations looks more complicated then the simple boost we used to start, it is the case for the parameters
we shall need to use. But it is of great use for the final simplifications and results.
Setting $y$ and $z$ to be solutions of :
\begin{eqnarray}
\sin(z)&=&\frac{-2\cos(z)\cosh(y)\sinh(x)}{k_1-k_2},\\
\sinh(y)&=&\frac{2\cosh(y)\cosh(x)}{k_1+k_2},\label{eqy} \\ \nonumber \\
\Rightarrow \tan(z)&=&\frac{-2\cosh(y)\sinh(x)}{k_1-k_2}  \nonumber \\
\Rightarrow \tanh(y)&=&\frac{-2\cosh(x)}{k_1+k_2} \nonumber
\end{eqnarray}
we have
\begin{eqnarray}
M_1''=\lp\ba{cccc}A & -\bar{M} & 0 & 0 \\ \bar{M} & -A & 0 & 0\\ 0 & 0 & B & \bar{M} \\
0 & 0 & -\bar{M} & -B\ea\rp,
\end{eqnarray}
with,
\begin{eqnarray}
A=\frac{1}{2}\lp(k_1-k_2)\cos(z)+(k_1+k_2-\frac{4\cosh(x)^2}{k_1+k_2})\cosh(y)\right. \nonumber \\
\left.+\frac{4\cos(z)\sinh^2(x)\cosh^2(y)}{k_1-k_2}\rp, \\
B=\frac{1}{2}\lp(-k_1+k_2)\cos(z)+(k_1+k_2-\frac{4\cosh(x)}{k_1+k_2})\cosh(y)\right. \nonumber \\
\left.-\frac{4\cos(z)\sinh^2(x)\cosh^2(y)}{k_1-k_2}\rp,\\
\bar{M}=\frac{\cosh(y)\sinh(2x)}{k_1+k_2}.
\end{eqnarray}
Taking a look at this $M_1''$ matrix we see it has the same form as the $M_0$ matrix so we shall diagonalise it using the same boost transformation. This time
the transformation will not be a boost since some parameters can be complex.
In fact before going any further we have to find the solution to equation (\ref{eqy}) that night be complex, $k_1+k_2$ is small
so the fraction on the right hand side is always bigger than one. We shall have to set $y=i\frac{\pi}{2}+\eta$, this little trick
allows us to easily solve all these equations. Finally to finish the diagonalisation we transform using the $v_1$ matrix :
\begin{eqnarray}
M_1'''&=&\tau v_1^T \tau M_1'' v_1 \nonumber \\
&=&\tau v_1^T\tau\tau v_2^T\tau M_1'v_2v_1 \nonumber \\
&=& \tau v_3^T\tau M_1'v_3,
\end{eqnarray}
with $v_3=v_2v_1$.
This last transformation can also be done using a slightly different matrix the $v_1'$ matrix defined by :
\begin{eqnarray}
v_1'&=&\lp\ba{cccc}\cosh(a) & \sinh(a) & 0 & 0 \\ \sinh(a) & \cosh(a) & 0 & 0 \\
0 & 0 & \cosh(b) & -\sinh(b) \\0 & 0 & -\sinh(b) & \cosh(b)\ea\rp, \\
v_1'^T\tau v_1'&=&\tau.
\end{eqnarray}
We finally have for the $M_1'''$ matrix the form
\begin{eqnarray}
M_1'''=\lp\ba{cccc}\xi_1 & 0 & 0 & 0 \\ 0 & -\xi_1 & 0 & 0 \\
0 & 0 & \xi_2 & 0 \\ 0 & 0 & 0 & -\xi_2 \ea\rp,
\end{eqnarray}
with
\begin{eqnarray}
\ba{c}\bar{M}\xi_1=(A^2-\bar{M}^2)\sinh(2a), \\
\bar{M}\xi_2=(B^2-\bar{M}^2)\sinh(2b),\ea
\end{eqnarray}
where
\begin{eqnarray}
\ba{c}\bar{M}\cosh[2a]=A \sinh(2a), \\
\bar{M}\cosh[2b]=B \sinh(2a).\ea
\end{eqnarray}
Using these transformation the diffusion matrix ${\mathcal V}$ can be expressed in terms of the diagonalisation matrices
in the way :
\begin{eqnarray}
{\mathcal V}=\tau(s_1s_2v_1')^T \tau E(s_1s_2v_1').\label{matrice1}
\end{eqnarray}
This form will be in fact far more simple then all the other possible ones, so this is the reason why we decided to use
it rather then the form with the scalar products of the eigenvectors. What we do is exactly the same since we work in
a base where the matrix of motion equations  is diagonal. If we keep the scalar products with the $\tau$ matrix
we see that the only to vectors having negative values are the negative moving ones, with a simple calculation we could link
the $\tau$ matrix to the helicity operator. This final expression we obtained will lead to simple results and is used to
compute all the reflection and transmission amplitude the results are given in the next section.
\subsubsection{Small sized Q balls}
\begin{figure}
\begin{center}
\includegraphics{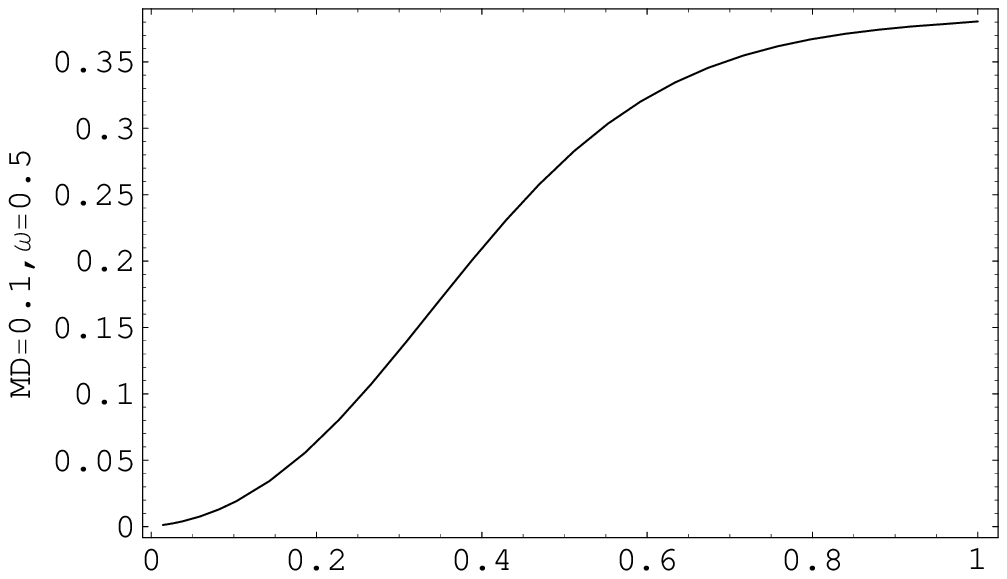}
\caption{Particle production rate for small values of the size parameter.}
\label{size}
\end{center}
\end{figure}

Using the results of previous section we where able to compute all the amplitudes for small sized Q balls. The method used was to replace the 
exponentials in the $E$ matrix by : $(1-ip_{1,2,3,4}l)$. These amplitudes still have complicated expressions but all of them exept $t_{22}$ are 
proportional to the size parameter $Ml\equiv l$. In the limit where $l$ goes to zero all the amplitudes fall to zero except  $t_{22}$ going to one.
The $t_{22}$ amplitude representing the probability of a fermion remaining a fermion. This probability is obviously one if the Q ball disappears.
If the size parameter is small the amplitudes will become proportional $l^2$ as for massless particle production. This quadratic behaviour
is shown on figure \ref{size}.
\section{Results of numerical integration}
\begin{figure}
\begin{center}
\includegraphics{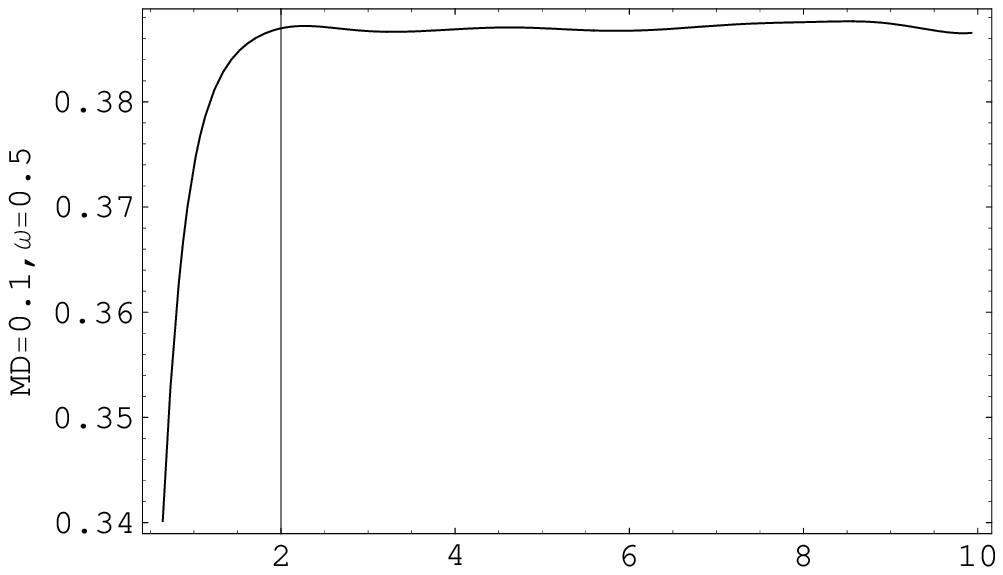}
\caption{Particle production rate for small values of the size parameter.}
\label{size1}
\end{center}
\end{figure}

\begin{figure}
\begin{center}
\includegraphics{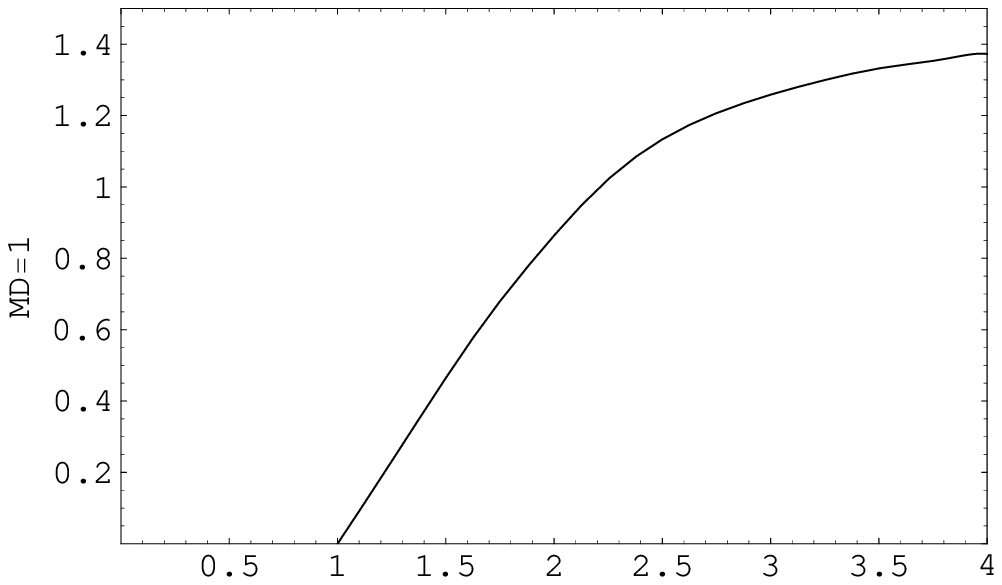}
\includegraphics{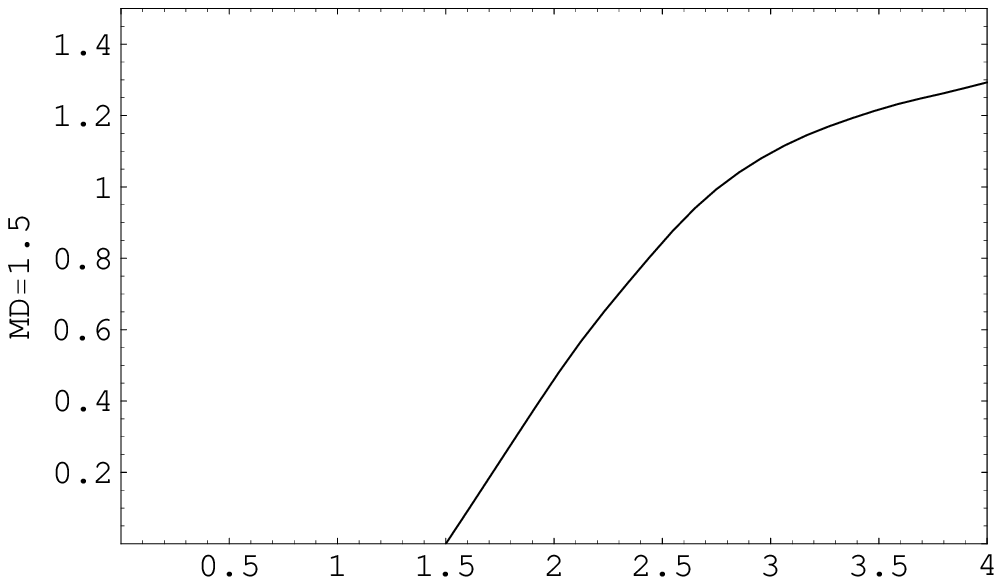}
\caption{Evaporation rate, $\frac{2\pi dN}{Mdt}$ for infinite (very big) Q-Balls in function of the frequency parameter for different values of the fermion 
mass parameter bigger then one.}
\label{big}
\end{center}
\end{figure}

We first tested the stability of production rate in function of size to see if like in the previous case the particle production rate
becomes constant and stable for big values of the size. If the production becomes constant above a certain size then we do not need to
care about complex averaging processes. Figure \ref{size1} shows the stability of evaporation rate for large Q balls.
The little oscillations are due to numerical instabilities that vanish for very big values of size. We can now compute the evaporation rate for 
values of the Dirac mass smaller then the Majorana mass (the coupling inside the Q ball).
\begin{figure}
\begin{center}
\includegraphics{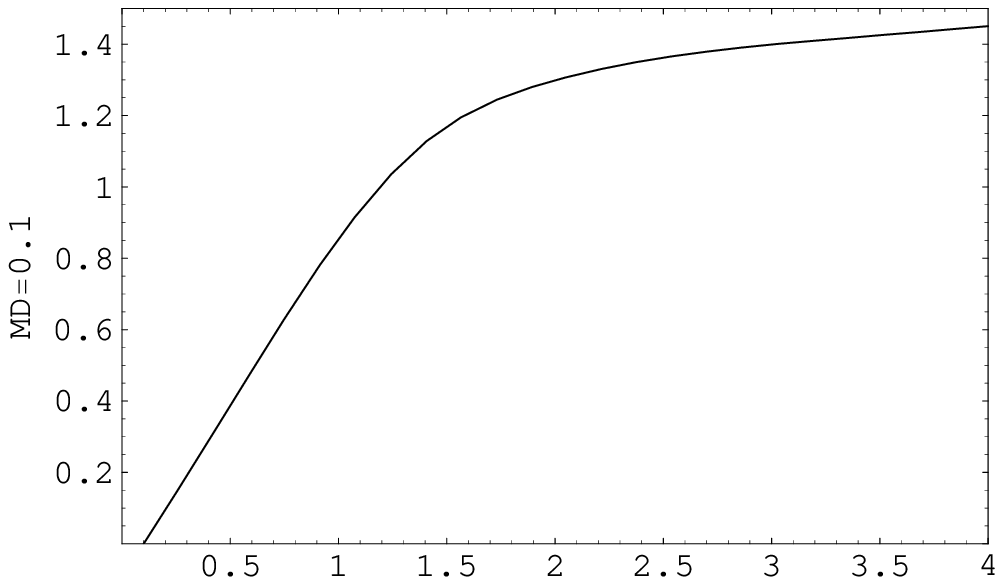}
\includegraphics{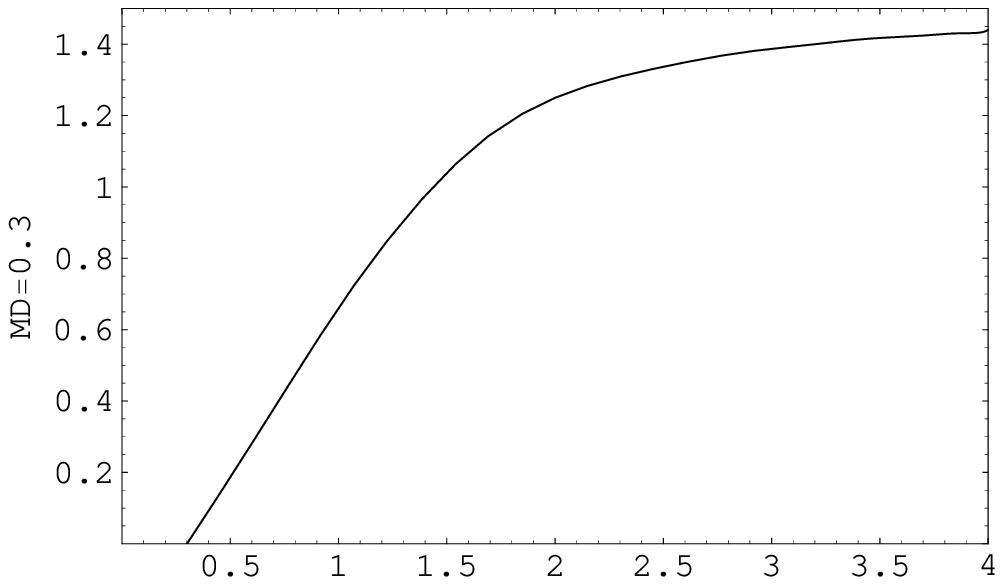}
\includegraphics{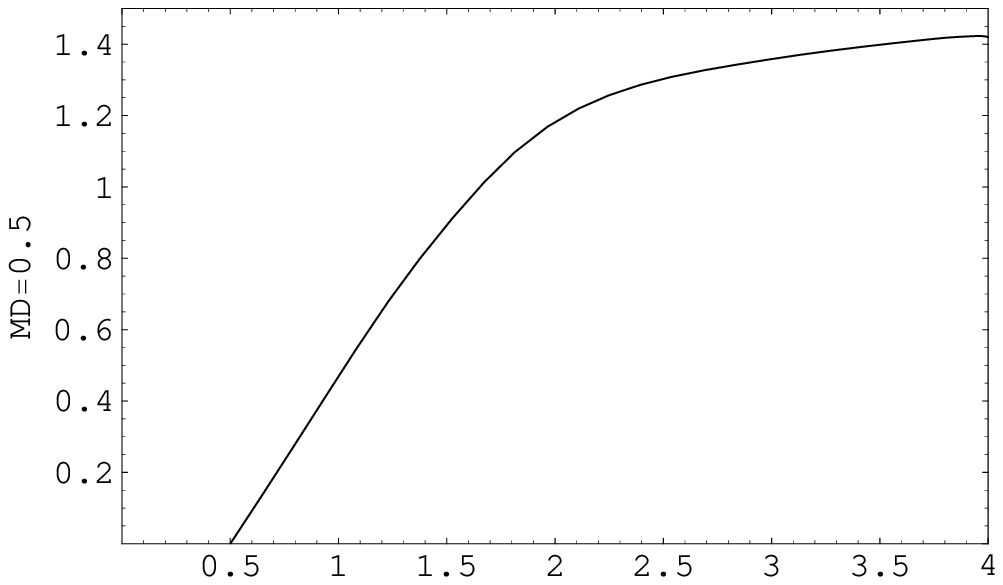}
\caption{Evaporation rate, $\frac{2\pi dN}{Mdt}$ for infinite (very big) Q-Balls in function of the frequency parameter for different values of the fermion 
mass parameter.}
\label{medium}
\end{center}
\end{figure} 
The next task we need to do is test the stability of our computations when the fermion mass parameter is bigger then the Majorana
coupling inside the Q-Ball, it is the case when $M_D\geq1$. This case shows exactly the same behaviour of the other one except for the
fact that it takes more computer time to obtain the plot of evaporation. The results are on figure \ref{big}. 
A quick analysis of these results shows that there is a superior limit for all parameter sets, this limit does not depend on
the mass parameter. Is seems to be normal since an infinite Q-Ball with an infinite internal frequency can produce any
mass fermions. The last words we shall say about these results is that the ``angle'' in the curve correspond to the value
of the frequency where the imaginary part of the impulsion inside the Q-Ball becomes zero, it is the point where particles
start to propagate inside the Q-Ball. The normalisation of evaporation by its upper bound will lead to the same shape as the massless case
but there will be a gap from zero to the value of the fermion mass.
\section{Energy flux far away from the Q ball}
The last step we need to achieve is compute the energy flux far away from the Q-Ball it is done by considering the flux through a sphere 
surrounding the Q ball. As before if the observer is far away from the Q ball the only important dimension is the distance to the Q ball.
We have,
\begin{eqnarray}
\frac{dE}{Mdtd\sigma}=\int_{-\frac{\om_0}{2M}+\frac{M_D}{M}}^{+\frac{\om_0}{2M}-\frac{M_D}{M}}\left(\frac{|r_{21}|^2+|t_{21}|^2}
{|r_{21}|^2+|r_{22}|^2+|t_{21}|^2+|t_{22}|^2}\right)\bar{\ep}^2d\bar{\ep},
\end{eqnarray}
the transmission amplitudes disappear when the Q-Ball's size is very big. This integration can be done numerically and we can also introduce the Q ball's
size to see its influence on the energy flux.
\begin{figure}
\begin{center}
\includegraphics{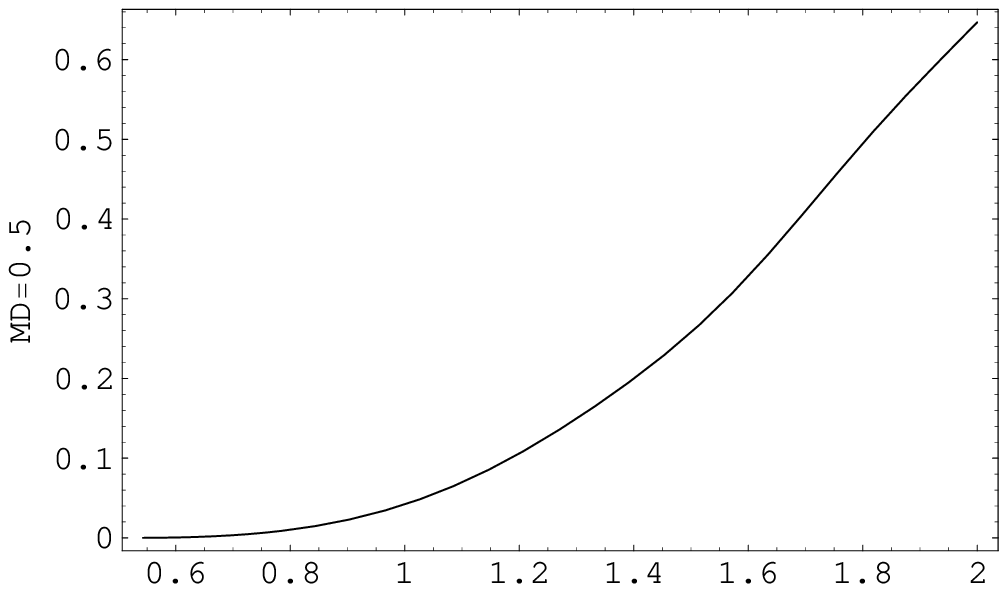}
\includegraphics{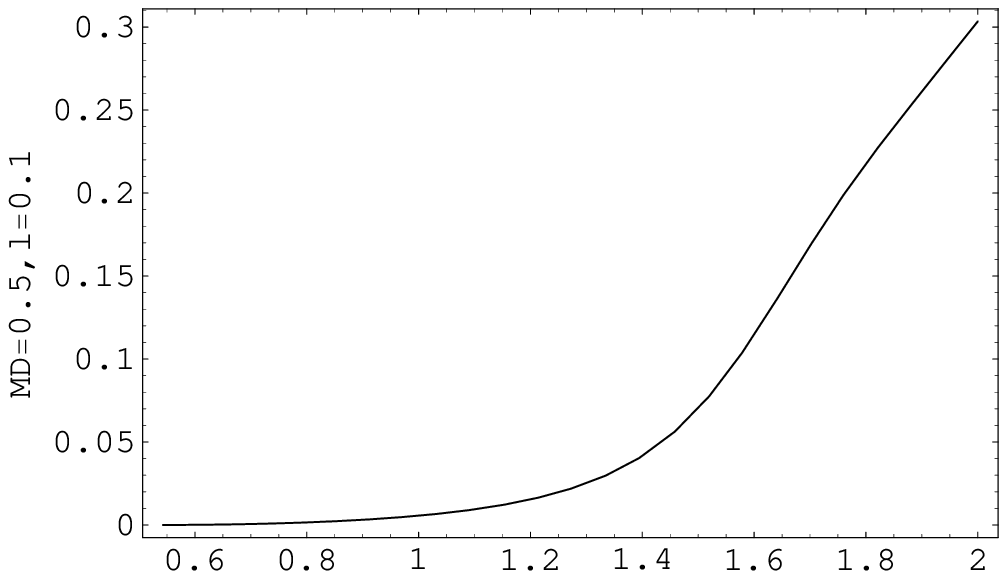}
\caption{Energy spectrum far away from the Q ball in function of $\frac{\om_0}{2M}$ once for a very big Q ball and once for a small one.}
\label{spec}
\end{center}
\end{figure} 
The only difference is that a small Q ball will produce less energy for until the value of the frequency parameter becomes big. We could normalise 
these figures with the absolute upper bound. This normalisation does not introduce any new features since the normalised curve for very big Q ball
would start with a constant part to then fall down. For the small one we will not find any constant part in the normalised curve.

\section{Conclusions}
As we have seen the coupling between the scalar field and fermionic field leads to particle production from the Q-ball \cite{Qballscohen}. To study this
particle production we constructed the exact quantum-mechanical state describing the particle producing Q-ball. We used Heisenberg's picture of quantum mechanics,
the state describing the produced fermions is fully characterised by the fact that no fluxes are moving towards the Q-ball. This condition is solved considering 
the asymptotics of the fields far away from the Q-ball. Using this state we constructed the Heisenberg field operator describing massless fermions produced 
by a Q-ball. This construction allowed us to prove that for large Q-balls in one space dimension the particle production does not depend on the Q-ball's size. 
While for small Q-balls the particle production rate is proportional to $l^2$. The extension of these results to three space dimensions is simple. 
For large Q-balls particle production is an evaporation, while for small Q-balls it depends on the size. The other result we need to point out is that we can
consider a variety of kinematical constructions to compute the evaporation rates. The first one is the standard one where we compute the reflection and 
transmission amplitudes for an incoming wave. The second one we used is based on the fact that no particles are moving towards the Q-ball. We proved that
these two pictures are equivalent.

The fact that fermions acquire a Dirac mass does not introduce many changes. In $1\oplus1$ dimensions the particle production rate does not depend on the
Q-ball's size for sufficiently big ones. This result is not very surprising, since taking the limit $m\rightarrow0$ leads to the same results as the coupling with 
massless fermions. In this case the only difference is that evaporation occurs in a different range. The internal frequency $\om_0$, the energy of one single
scalar forming the Q-ball, must be bigger than the produced fermion mass. This result is also quite intuitive, the scalars forming the Q-ball desintegrate into
fermions so their energy must be bigger than the fermion mass. The second fact is that particle production occurs in the range mixing positive and negative
frequency terms. In this range the Bogolubov transformations we build are non trivial. Using these two results we proved that evaporation can only take
place in the range : $[M_D-\frac{\om_0}{2};\frac{\om_0}{2}-M_D]$, with $\frac{\om_0}{2}\geq M_D$. This result is in total accordance with the previous
work done on the subject \cite{Qballscohen,Qballsnew}, and extends it a significant way. This new definition for the evaporation range will introduce
a new upper bound for the evaporation rate.

When the Q ball's size becomes small there is no more evaporation since the production rate depends on the size. For small sized Q balls the particle
production rate is proportional to $l^2$. The size will also slow down  the energy flux a distant observer can measure. Taking the limit 
$l\rightarrow\infty$ does not need any complex averaging processes since the evaporation rate is constant and $l$ independent for big values of the size.

We also computed all the transmission and reflection coefficients for a massive fermion being scattered by a large Q-ball. This construction allowed us to
compute the exact profile of the evaporation rate. Using these profiles we proved that both constructions are equivalent. The last result we have proved
is that evaporation rate is proportional to $\om_0$ in one space dimension while it is proportional to $\om_0^3$ in three dimensions. In fact in three
dimensions it is proportional to $(\frac{\om_0}{2}-M_D)^3$. These reflection and transmission amplitudes will be used to study the behaviour of Q Balls
in matter.

\begin{acknowledgments}
The author would like to thank M. Shaposhnikov for suggesting this problem and for useful discussions.
\end{acknowledgments}

\end{document}